\documentclass[lettersize,journal]{IEEEtran}
\usepackage[T1]{fontenc}
\usepackage{amsmath,amssymb}
\usepackage{graphicx}
\usepackage{paralist}
\usepackage[inline]{enumitem}
\usepackage{listings}
\usepackage{algorithm}
\usepackage{hyperref}
\usepackage{xcolor}
\usepackage{macros}
\usepackage{todonotes}
\usepackage{ulem}
\usepackage{orcidlink}
\usepackage{balance}

\newcommand{\labels}{\ensuremath{\mathsf{alph}}}

\definecolor{keyworkColor}{rgb}{0.5,0,0.35}
\definecolor{stringColor}{rgb}{0.58,0,0.82}

\definecolor{keywordcolor}{rgb}{0.0, 0.5, 0.0} 
\definecolor{stringcolor}{rgb}{0.0, 0.0, 1.0}  
\definecolor{commentcolor}{rgb}{0.5, 0.5, 0.5} 

\newtheorem{definition}{Definition}
\newtheorem{lemma}{Lemma}
\begin{document}

 \title{BEST: A Unified \uline{B}usiness Process \uline{E}nactment 
 via \uline{S}treams and \uline{T}ables for Service Computing 
}

%


\author{\IEEEauthorblockN{Ahmed Awad\IEEEauthorrefmark{1},\IEEEauthorrefmark{2}\orcidlink{0000-0003-1879-1026},Feras Awaysheh\IEEEauthorrefmark{3},\IEEEauthorrefmark{5}\orcidlink{0000-0002-9561-6099}, and Hugo A. López\IEEEauthorrefmark{4}\orcidlink{0000-0001-5162-7936}}
\\
 \IEEEauthorblockA{\IEEEauthorrefmark{1}The British University in Dubai, Dubai, The UAE
 Email: ahmed.awad@buid.ac.ae}\\
 \IEEEauthorblockA{\IEEEauthorrefmark{2}Cairo University, Giza, Egypt}\\
 \IEEEauthorblockA{\IEEEauthorrefmark{3}Tartu University, Tartu, Estonia
 Email: feras.awaysheh@ut.ee}\\
\IEEEauthorblockA{\IEEEauthorrefmark{5}Department of Computing Science, Umeå University, Umeå, Sweden
 Email: feras.awaysheh@umu.se}\\
 \IEEEauthorblockA{\IEEEauthorrefmark{4}Technical University of Denmark, Kgs. Lyngby, Denmark
 Email: hulo@dtu.dk}

%

}

%
%

\maketitle
\begin{abstract}


Business process models are essential for the representation, analysis, and execution of organizational processes, serving as orchestration blueprints while relying on (web) services to implement individual tasks. At the representation level, there are two dominant paradigms: procedural (imperative) notations that specify the sequential flows within a process and declarative notations that capture the process as a set of constraints. Although each notation offers distinct advantages in representational clarity and cognitive effectiveness, they are seldom integrated, leading to compatibility challenges. In this paper, we set aside the imperative-declarative dichotomy to focus on orchestrating services that execute the underlying tasks. We propose an execution semantics based on the Continuous Query Language (CQL), where CQL statements respond dynamically to streams of events. As events unfold, these CQL statements update the execution state (tables) and can generate new events, effectively triggering (web) services that implement specific process tasks. By defining all executions around a unified event model, we achieve cross-language and cross-paradigm process enactment. We showcase how industrial process modeling languages, such as BPMN and DCR graphs, can be enacted through CQL queries, allowing seamless orchestration and execution of services across diverse modeling paradigms.

\end{abstract}

\section{Introduction}
Business processes are at the core of successful digital transformation initiatives. Process mining has proven valuable as a data-driven process analytics and enhancement approach~\cite{PMBook16}. Through process mining, process models can be discovered, deviations between models and the actual execution can be identified~\cite{conformanceChecking2018}, and improvements can be suggested~\cite{WICKRAMANAYAKE2023106678,DBLP:conf/bpm/SeeligerSM17}. However, the utmost value of these insights will not materialize unless there is a solid and flexible process execution (enactment) infrastructure that can reflect the changes in running or future process instances. 

The execution infrastructure can be further sub-divided into atomic services, e.g. placing an order, and service compositions that logically relate and orchestrate the invocation of several other (atomic) services to fulfill a business goal. Service-oriented architecture and the recent micro-services architecture have contributed the flexibility of service execution by, among other things, modularizing and localizing the effect of change and failure. For service composition, business process models are generally used~\cite{buchwald2012bridging, 7387773}.

Generally, processes can be modeled following procedural, or declarative paradigms~\cite{DBLP:conf/bpm/FahlandMRWWZ09}, where procedural dictates only allowed behavior, and declarative defines the scope of the process, allowing multiple types of executions that do not violate the specification. While procedural languages fit better well-defined processes, e.g. loan applications, declarative languages provide flexibility to knowledge-driven processes, e.g. healthcare~\cite{coopis24}. Moreover, some processes may have standardized and knowledge-intensive parts, benefiting from hybrid modeling languages~\cite{mixedParadigms2013,DBLP:conf/otm/SlaatsSMR16}.

When implementing process models, its execution infrastructure consists of several information systems, external services, and human performers interacting to complete a business process instance. A process execution engine usually orchestrates the interaction and the order of execution of the different activities of a business process. There have been several technologies and solutions for process enactment and orchestration from academia and industry. YAWL~\cite{YAWL} is a prominent open-source workflow engine. BPEL~\cite{BPEL} is a standard for enacting business processes using web services. BPMN~\cite{omg2014bpmn} is a more recent and widely accepted standard for modeling and enacting procedural business processes that are supported by several engines like Camunda~\footnote{\label{footnote:camunda}\url{https://camunda.com/}}, Bizagi~\footnote{\label{footnote:bizagi}\url{https://www.bizagi.com/}}, Activiti~\footnote{\url{https://www.activiti.org/}}, and many more. DCR Graphs~\cite{hildebrandt2011declarative} is currently the only declarative process modeling notation supported by industrial players, with both commercial~\footnote{\url{https://www.dcrgraphs.net/}} and open source implementations~\cite{dcr4py}. Other declarative notations, for example, Declare~\cite{DBLP:conf/edoc/PesicSA07}, can be defined and executed via encodings to Coloured Petri Nets~\footnote{\url{https://www.win.tue.nl/declare/}}.  As these engines vary in their support for the execution semantics of the different modeling constructs~\cite{DBLP:journals/fgcs/GeigerHLW18}, migrating running instances to a changed process model or deploying newer versions of the orchestrator has considerable technical debt. Moreover, despite the visible advantages of hybrid process models, none of the existing engines have adopted hybrid process execution.

This paper introduces a novel approach, BEST, to enacting and orchestrating business processes. We argue that by using streams, tables~\cite{DBLP:conf/birte/SaxWWF18}, and the continuous query language (CQL)~\cite{DBLP:conf/dbpl/ArasuBW03}, we can implement the execution semantics of several business process notations. This complements the flexibility in service-oriented architecture by bringing flexibility to the orchestration level. Moreover, as we adopt the streams and tables abstraction, we are able to enact hybrid process models and service compositions. 

This paper will show the feasibility of our approach by instantiating both imperative BPMN models and declarative DCR models in our framework. 

Our contributions to this paper are:
\begin{enumerate}
    \item An architecture where a CQL-compliant stream processing engine can serve as a process orchestrator,
    \item A rule-based approach to express the common modelling constructs as CQL statements,
    \item An implementation based on Esper,
    \item A discussion about the flexibility this approach brings in terms of supporting the evolution of process models and instance migration and the support of executing mixed (hybrid) process models.
\end{enumerate}

The rest of this paper is organized as follows: Sec.~\ref{sec:preliminaries} presents preliminary concepts necessary to follow the rest of the paper. The main contribution of the paper is presented in Sec.~\ref{sec:declarative:orchestration}. The implementation and evaluation details are discussed in Sec.~\ref{sec:implementation}. Related work is discussed in Sec.~\ref{sec:related:work} before we conclude the paper in Sec.~\ref{sec:conclusion}.








\section{Preliminaries}
\label{sec:preliminaries}


This section provides the necessary background to understand and position the contributions in this paper. Section~\ref{sec:process:models} provides the formalization of process models, capturing the commonalities between procedural and declarative process models regarding their execution. The streams and tables duality from the stream data analytics domain and the continuous query language are discussed with sufficient detail in Section~\ref{sec:stream:table:duality}. 

\subsection{Process Models}
\label{sec:process:models}

While the literature acknowledges the differences between imperative and declarative process modeling notations \cite{DBLP:conf/caise/FahlandLMRWWZ09}, we decided to focus on its similarities. Any (imperative or declarative) formalism with trace-based semantics can be used as basic processes, and we have no requirements that all underlying processes
are specified in the same notation. 

We abstract the underlying formalism into
the notion of a (abstract) \emph{process notation}~\cite{DBLP:conf/ifm/DeboisLSAH20}. 
Assume a fixed universe $\cal{U}$ of actions. 

\begin{definition}[adopted from~\cite{DBLP:conf/ifm/DeboisLSAH20}]
  \label{def:abstract}
  A \emph{process notation} $A= \langle{\cal P}, \labels, \excluded, $ $\step\rangle$ comprises a set
  ${\cal P}$ of process models;  a labelling function $\mathsf{alph} : {\cal P} \to 2^{\cal U}$, 
  an exclusion function $\excluded : {\cal P}\to 2^{\cal U}$; 
  and a transition predicate  $\step : {\cal P} \times
  {\cal U} \times {\cal P}$. We require that $\langle P,l,Q\rangle  \in \step$ implies both
  $l \in \alphabet P$ and $\alphabet P = \alphabet{Q}$, and if also $(P,l,Q') \in \step$
  then $Q=Q'$, that is, $\step$ is \emph{action-deterministic}.
\end{definition}
Intuitively, $\mathsf{alph}$ gives a finite bound on the actions a process may exhibit, 
and we require this bound to be preserved by $\step$-transitions. Similarly, $\excluded$
tells us which actions are excluded in a given process; this set can change as
the process evolves.

Notice that for most imperative notations, an instantiation of $A$ with $\excluded=\emptyset$ is sufficient. We provide the example of BPMN:

\begin{definition}[BPMN (adapted from~\cite{DBLP:journals/infsof/GorpD13})]
\label{def:bpmn}
A BPMN model is a tuple $\langle\mathcal{F}_e, \mathcal{F}_{id}, \mathcal{F}^{id}_{e}, \mathcal{D}, \mathcal{V},\mathcal{S}_f, type, exp\rangle$, where:
\begin{enumerate*}
    \item $\mathcal{F}_e \subseteq \mathcal{T} \cup \mathcal{E} \cup \mathcal{G}$ is a finite set of flow elements consisting of tasks, events, and gateways, which are pairwise disjoint,
    \item $\mathcal{F}_{id}$ is a finite set of flow elements identifiers,
    \item $\mathcal{F}^{id}_{e}: \mathcal{F}_{e} \rightarrow \mathcal{F}^{id}$ assigns each flow element a distinct identifier,
    \item $\mathcal{D}$ is a finite set of data objects, also known as case variables,
    \item $\mathcal{V}$ is a finite set of values,
    \item $\mathcal{S}_f \subseteq \mathcal{F}_e \times \mathcal{F}_e$ is a finite set of sequence flow edges,
    \item $type: \mathcal{E} \cup~\mathcal{G} \rightarrow \{start, intermediate, end, AND, XOR, OR\}$ is a function that defines further types for events and gateways,
    \item $exp: \mathcal{S}_f \rightarrow EXP(\mathcal{D},\mathcal{V}) \cup \{true\}$ is a function that assigns a Boolean expression to the sequence flow edges.
\end{enumerate*}
\end{definition}

To describe the transition relation $\emph{step}$, we can use any encoding of the language into a labeled transition system. For instance \cite{KRISHNA20191} provides an LTS for the set of workflow operators in BPMN, and \cite{corradini2016operational} provides an LTS for BPMN collaboration diagrams. For the sake of space, we assume that such encoding exists, and recall \cite{KRISHNA20191} for details on its construction.

\begin{lemma}
Let a BPMN model be the tuple $BPMN =\langle\mathcal{F}_e, \mathcal{F}_{id}, \mathcal{F}^{id}_{e}, \mathcal{D}, \mathcal{V},\mathcal{S}_f, $ $type, exp\rangle$ according to Definition~\ref{def:bpmn}. Let ${\cal P}$ be the set of BPMN models, and let the labeling function be defined as $\lambda \colon BPMN.\mathcal{F}^{id}_e$, and let $\excluded: {\cal P}\to \emptyset$. Finally, we take $(P, t, P') \in \step$ if $t$ is a transition enabled in $P$ according to \cite{KRISHNA20191}. Then $A=\langle{\cal P}, \lambda, \excluded, \step\rangle$ is a process notation.
\end{lemma}

        \small \begin{definition}[DCR graphs \cite{hildebrandt2011declarative}]\label{def:DCR:graphs}
            A Dynamic Condition Response Graph (DCR Graph)  is a tuple $G = \langle \Events, \MarkingMatrx, \Acts, \conditionRel, \responseRel, \pm, l \rangle$, where \begin{enumerate*}
            \item $\Events$ is the set of events,
            \item $M \in \mathcal{M}(G)= \mathcal{P}(\Events) \times \mathcal{P}(\Events)$ $ \times \mathcal{P}(\Events)$ is a marking and $\mathcal{M}(G)$ is the set of all markings,
            \item $\Acts $ is the set of actions,
            \item $\conditionRel \subseteq \Events \times \Events$ is the condition relation,
            \item $\responseRel \subseteq \Events \times \Events$ is the response relation, 
            \item $\pm \colon \Events \times \Events \rightharpoonup \{ +, \% \}$ defines the dynamic inclusion/exclusion relations by $\Event \inclusionRel \Event^{\prime}$ if $\pm(\Event,\Event^{\prime}) = +$ and $\Event \exclusionRel   \Event^{\prime}$ if $\pm (\Event,\Event^{\prime})= \%$,
            \item $l \colon \Events \rightarrow \Acts$ is a labeling function.
            \end{enumerate*} 
        \end{definition} 

        In DCR graphs, the condition and response relations allow for cyclic interactions. The marking $M = (Ex, Re, In) \in \mathcal{M}(G)$ comprises three sets of events: executed events ($Ex$), pending responses ($Re$) that are yet to be executed or excluded, and currently included events ($In$). An event $e$ \emph{is enabled} in a marking $M= (\Executed, \Pending, \Included)$ if:
        \begin{enumerate*}[label= \arabic*)]
            \item $e \in \Included$ and 
            \item if $\exists f. (f,e) \in \conditionRel \implies f \in \Executed \lor f \notin \Included$.
        \end{enumerate*}


\small \begin{definition}[DCR Transitions             \cite{hildebrandt2011declarative}]   \label{def:DCRLTS}         
            For a DCR graph $G = \langle \Events, M, \Acts, \rightarrow\!\!\bullet,  \bullet\!\!\rightarrow, \pm, l \rangle$, the corresponding LTS $T(G)$ is the tuple $\langle \mathcal{M}(G), \MarkingMatrx, \rightarrow \subseteq \mathcal{M}(G) \times \Acts \times \mathcal{M}(G) \rangle$  where $\mathcal{M}(G)$ is the set of markings $G$, $\MarkingMatrx \in \mathcal{M}(G)$ is the initial marking, and $\rightarrow \subseteq \mathcal{M}(G) \times (\Events \times \Acts) \times \mathcal{M}(G)$ is the transition relation given 
            \EventTransition{M'}{(\Event,\Action)}{M''}
            where:
            \begin{enumerate*}
                \item $M^\prime=(Ex^\prime,Re^\prime,In^\prime)$ is the marking before the transition,
                \item $M^{\prime\prime}=(Ex^{\prime}\cup \{\Event\},Re^{\prime\prime},In^{\prime\prime})$ is the marking after the transition,
                \item $e \in In^\prime$,
                \item $l(\Event) = a$,  
                \item $\{ \Event^\prime \in In^\prime ~|~ \Event^\prime \rightarrow\!\!\bullet \Event \} \subseteq Ex^\prime$,
                \item $In^{\prime\prime} = (In^\prime ~\cup~ \{ \Event^\prime ~|~ \Event \rightarrow\!\!+ \Event^\prime \} ) \backslash \{ \Event^\prime ~|~ \Event \rightarrow\!\!\% \Event^\prime \}$, and
                \item $Re^{\prime\prime} = ( Re^\prime \backslash \{ \Event \} ) \cup \{ \Event^\prime ~|~ \Event \bullet\!\!\rightarrow \Event^\prime \}$.
            \end{enumerate*}            
        \end{definition} 

\begin{lemma}[\cite{DBLP:conf/ifm/DeboisLSAH20}]
  Take ${\cal P}$ to be the set of DCR graphs with labels in ${\cal U}$
  and injective labeling functions.  
  Let $\excluded$ be the function which given a DCR graph $G$ with events
  $\Events$, marking $\MarkingMatrx$, and labeling $l$ returns the set of labels of 
  events of $\Events$ that are not in $In$, that is, $\excluded G = \{ l(e) \mid 
  e \in E \setminus In \}$. Finally take $(G, l, G') \in \step$ iff
  there exists some event $e \in E$ s.t.~$\ell(e) = l$ and 
  $\transition{G}{e}{G'}$. Then $\langle{\cal P}, l, \excluded, \step\rangle$ is a process 
  notation. 
\end{lemma}

\subsection{Streams and Tables}
\label{sec:stream:table:duality}

A data stream is a potentially unbounded sequence of data items. Examples are ubiquitous. Updates of temperature sensors, stock market prices, and process execution events as known from the process mining field are all examples of such streams. 
Traditionally, two broad classes of systems process such data streams~\cite{DBLP:journals/is/AwadTLKVS22}. The first category is concerned with SQL-like operations on data streams. That is, computing aggregations, filtering, joining, etc. on one or more data streams. The other category is concerned with detecting patterns among sequences of stream elements, known as complex event recognition~\cite{DBLP:journals/vldb/GiatrakosAADG20}. In this case, data elements are first recognized as events. That is, they are instances of a data schema and are timestamped. An event pattern is similar to a regular expression. Once matched, we can derive a so-called complex event. For example, we can derive a fire event in some location if we observe a continuous increase in temperature simultaneously with a drop of humidity within the same period, e.g., five minutes.

Complex event recognition has been utilized to monitor the compliance of running process instances to regulations, best practices, and other forms of policies~\cite{DBLP:conf/sac/AwadBESAS15}, and for event correlation~\cite{DBLP:journals/is/HelalA22}. 

\begin{figure}
    \centering
    \includegraphics[width=0.5\linewidth]{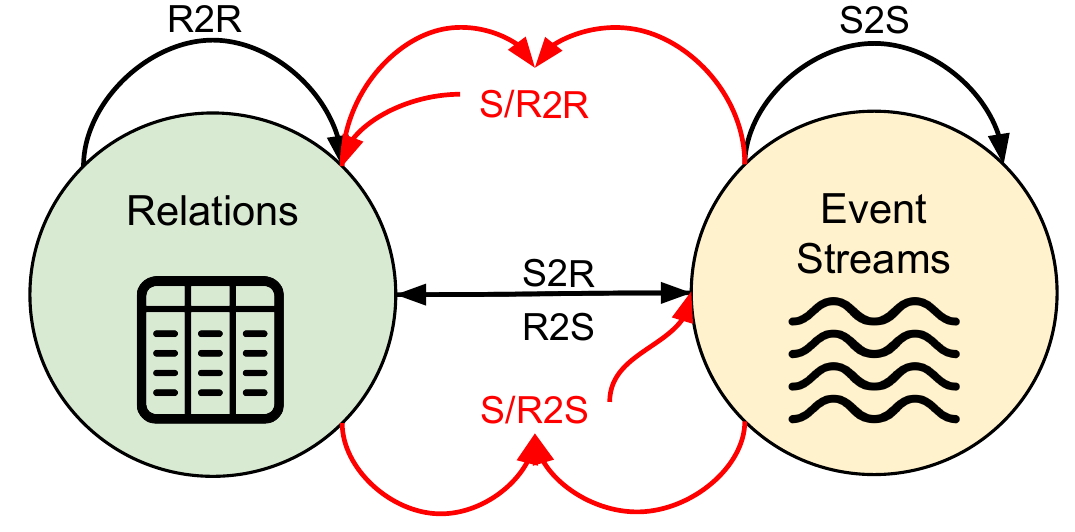}
    \caption{CQL conceptual model}
    \label{fig:CQL:model}
\end{figure}

A conceptual model that underpins the two classes of data stream processing was developed by Arasu et al.~\cite{DBLP:conf/dbpl/ArasuBW03,DBLP:journals/vldb/ArasuBW06} in their seminal work on the continuous query language extends the relational algebra by the notion of a stream in addition to the notion of relations (tables). Thereupon, four families of operators transform the data (see Figure~\ref{fig:CQL:model}) from 1) streams to relations (S2R), 2) streams to streams (S2S), 3) relations to relations (R2R), and 4) relation to streams (R2S). 

All the standard SQL operators fall under the (R2R) family. Any conventional SQL statement takes one or more relations as input and results in another relation. S2S operators can further be categorized into stateless and stateful operators. The former involves transformations like projection, i.e., selecting a subset of the attributes of the input data. The latter includes Complex Event Recognition (CER) pattern operators~\cite{DBLP:conf/debs/ArtikisMUVW17}, stream-to-stream joins~\cite{DBLP:conf/edbt/JafarpourD19}. S2R operators are commonly known as window operators~\cite{verwiebe2023survey}. One or more data elements are grouped following grouping criteria. Mostly, stream elements are grouped within time windows. For example, group temperature sensor readings every five minutes. In this case, temperature readings from the same sensor are added to a flushed and refilled buffer every five minutes. Elements can be windowed by other means than time. For a comprehensive discussion of window types, we refer the reader to the survey in~\cite{verwiebe2023survey}. The inverse family of operators R2S is used to stream out whatever processing logic is applied to the content of the windows. Referring to the temperature example, the content of the five-minute window can be averaged where the average is emitted on a stream that can later on be used to update some monitoring dashboard.

We can observe that the four families are homogeneous like their inputs. That is, the inputs are either streams or relations. However, new family operators have recently been developed to cover new needs. For example, it is a common practice to perform so-called data enrichment. That is, a stream data element gets enriched with more details that are usually residing in a relation. This is called stream-relation join. The result of such join can be used to update another relation, in this case, S/R2R, or the enriched data is emitted on another stream for further downstream processing, S/R2S in this case. The families of S2R, R2S, S/R2R, and S/R2S contribute to the stream-table duality~\cite{DBLP:conf/birte/SaxWWF18}. That is, elements on the stream can update an append-only table (relation); therefore, we can get a snapshot view of the current state at any point in time. Inversely, when scanned in chronological order or their arrival, the records in the table can replay the stream.

In the context of this paper, we leverage the stream-table duality operators to build a process execution engine by defining what data streams and tables are needed to allow process instances orchestration respecting control and data flow specifications in process models.

\subsection{Process Execution}

To ease the follow-up, we define the basic inputs common for process execution and other process analytics tasks, i.e., process events. Next, we discuss concrete examples of the families of the CQL operators discussed above. 

It is widely accepted that events are the atomic entities that reflect the progress of process execution. Raw execution events, Definition~\ref{def:raw:event}, are generated to manifest the execution status of process steps. The events are generated following a lifecycle model. In the context of this paper, we follow a simple three-state model of \texttt{started}, \texttt{completed}, or \texttt{skipped}.
\begin{definition}[Raw Execution Event]\label{def:raw:event} A raw execution event is a tuple $re = \langle model, case, node, state, payload, executor, ts\rangle$ where $model$ refers to the process model, $case$ is the case identifier, $node$ is the node in the process model for whose update the event was generated, $state \in \{started, completed, skipped\}$ is the lifecycle state for the node, $payload$ is the data payload of the event, $executor$ refers to the entity that performed the $node$ of the event, and $ts$ is the timestamp at which the event was generated.  
\end{definition}

Attributes $case$, $node$, $ts$, and $payload$ are straightforward and are common in business process management literature. The attribute $model$ refers to the process model definition from which the instance $case$ was instantiated. This is important to track structural changes in process models and to correlate execution events with the process model version correctly. Section~\ref{sec:flexibility} will discuss this in more detail. The $node$ attribute refers to any identifiable node in the process model. Tasks, decision points, synchronization points, and start and end points are all nodes. Nodes can be further classified depending on the context and the process modeling language. For instance, in BPMN, a node can be classified as an Event, Gateway, Activity, etc (Definition~\ref{def:bpmn}). $state$ refers to the execution state of the node and is usually assigned following a task lifecycle model. In the context of this paper and without loss of generality, we stick to the states $started, completed,~and~skipped$. Particularly, the $skipped$ state plays a crucial role in the orchestration and the synchronization decision of join nodes, e.g., AND or OR joins in BPMN.

$Payload$ relates to contextual data of the process instance and refers to the data content of the event that can be used, e.g., in decision points. The $payload$ attribute should be used to update instance-specific global variables that can be used to activate certain paths in the process and to be passed to future tasks. The $executor$ refers to the entity that performed the $node$ of the event. This can be a reference to a human performer or a software agent. However, for the rest of this paper, we will not consider the $executor$ attribute as we focus on the control and data flow aspects of process execution.

For simplicity, in the rest of the paper, to access a component of an event, we will use the dot notation: given an event $e = (m, c, n, s, p, ex, t)$, we can obtain, e.g., the value of $m$ by using $e.m$.

\begin{definition}[Event stream]\label{def:event:stream} Let $E$ be the set of 
all events. An event stream $\mathcal{S} : \mathbb{N} \to E$ is a possibly unbounded sequence $\sigma=\langle e_1,e_2, 
\dots \rangle$, such that $e_x.ts \leq e_y.ts$, where $x < y$.
\end{definition}

The Event Stream is a raw event stream fed by both the execution engine (orchestrator) and external task executors; more on this in Sec.~\ref{sec:overview}. These events can be accumulated and aggregated as tables or windows over the stream. 


\begin{definition}[(State) Table]\label{def:event:window} Let $S$ be an event stream, let $\mathrm{T}$ be the set of timestamps, and $Lc =\{started, completed, skipped\}$ a set of lifecycle states. A state table $\mathcal{R}$ is defined as $\mathcal{R} \subseteq \mathcal{M} \times \mathcal{C} \times \mathcal{N}\times Lc \times \mathrm{T}$ whose primary key $PK \subseteq \mathcal{M} \times \mathcal{C} \times \mathcal{N}$, where $\mathcal{M}$ is the universe of process models, $\mathcal{C}$ is the universe of case identifiers, and $\mathcal{N}$ is the universe of nodes identifiers within process models. The arrival of a new event $e \in \mathcal{S}$ may result in an upsert operation into $\mathcal{R}$.
\end{definition}

The state table $\mathcal{R}$ represents the relations part in the CQL model, c.f. Figure~\ref{fig:CQL:model}. However, there can be more than one table, not necessarily the dual of the event stream, to support process execution. For example, there can be tables that track case variables, tables that track human resources, a table that tracks the cases, process models, etc. As stated in Definition~\ref{def:event:window}, the update or insert (upsert) operation on the table is a reaction to the arrival of new events on the stream. The reaction is defined as a CQL statement that is triggered by the arrival of a new event. In the following, we give general examples of how such reactions can be defined as CQL statements before we discuss the use of streams, tables, and CQL for process orchestration.

CQL operators can react to events arriving on the Event Stream in several ways. Let us consider the following requirements:
\begin{itemize}
    \item R1: Alert for the execution of tasks that are rarely executed in processes, 
    \item R2: Keep track of tasks that have been skipped, 
    \item R3: Update the number of newly created process instances every five minutes, 
    \item R4: Track the number of newly created process instances by their process model version 
    
\end{itemize}

\begin{lstlisting}[language=SQL,label={sql:example:CQL}, caption={CQL statements for Requirements R1-R4}]
-- R1 an S2S operator
insert into RarelyExecutedTasksStream
select * from EventStream where node='RareTask';
-- R2 S2R operator mapping one event at a time
insert into SkippedTasksTable
select * from EventStream where state='skipped';
-- R3 S2R operator mapping multiple records and R2S 
insert to NewInstancesSummary --R2S
select count(case) from EventStream.window:time(5 min) --S2R
where node='start';

-- R4 an S/R2S operator
insert to NewInstancesByProcessVersionStream 
select E.case, P.version  from EventStream as E join ProcessModelTable as P 
on E.m = P.id
where E.node='start';

\end{lstlisting}

Listing~\ref{sql:example:CQL} shows the implementation of the requirements above in CQL, using the EPL syntax. To differentiate between streams and tables, we use a naming convention that attaches either the suffix `Stream' or `Table.' The first CQL statement is an $S2S$ operator that maps events on the \texttt{Event Stream} to events on the \texttt{Rarely Executed Tasks Stream}. The events on the former stream are filtered following the name of the task (node) that is known to be part of the rarely executed portion of the process model. The second CQL statement, $S2R$, adds skipped events observed on the \texttt{Event Stream} to the \texttt{Skipped Tasks Table}. Each matching event on the stream will correspond to one record in the table. The third CQL statement is another $S2R$ operator that maps many events on the stream to a table, performs an aggregation, and then sends the aggregate value on another stream, $R2S$. The intermediate table is computed on the fly by the collected events in the five-minute time window. The last CQL statement represents a stream-table join. Each arriving event is enriched with information from the \texttt{Process Model Table}; the case identifier and the process model versions are emitted on the \texttt{New Instances By Process Version Stream}. In Section~\ref{sec:declarative:orchestration}, we will show how we can use streams, tables, and CQL operators to build a process execution engine.




\section{Declarative Orchestration for Process Models}
\label{sec:declarative:orchestration}

\begin{figure}[t]
    \centering
    \includegraphics[width=1\linewidth]{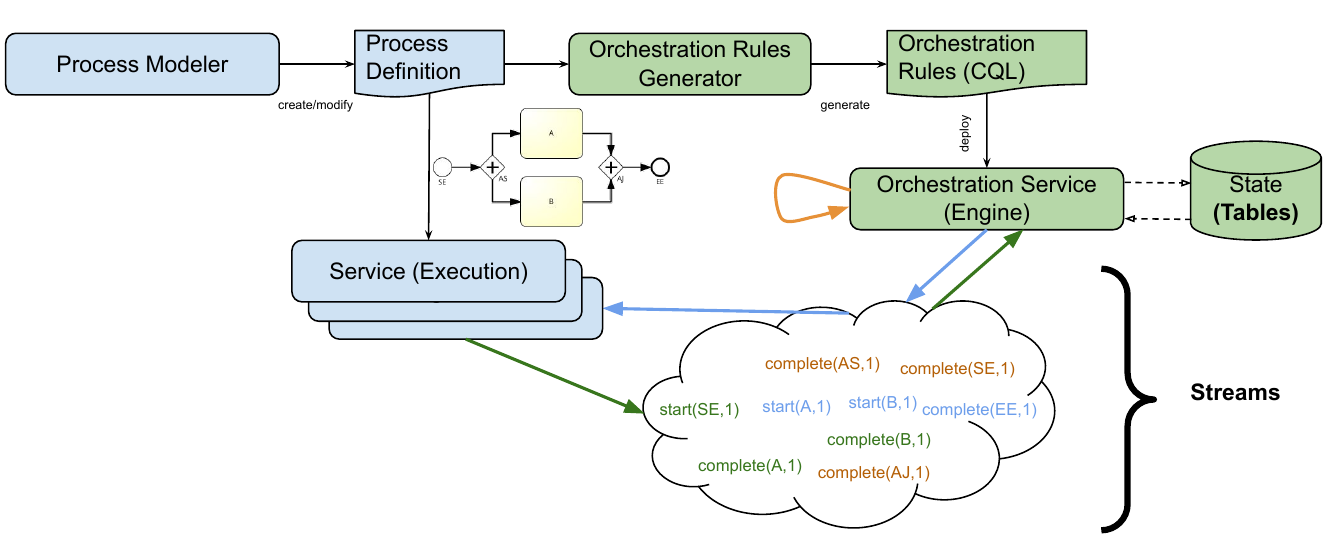}
    \caption{Architectural view, components colored in green are the focus of this paper}
    \label{fig:architecture}
\end{figure}


\subsection{Overview}
\label{sec:overview}

\begin{figure}[t]
    \centering
    \includegraphics[width=1\linewidth]{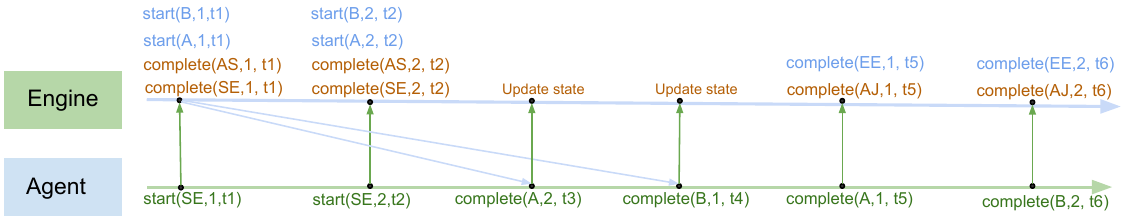}
    \caption{Example execution}
    \label{fig:execution}
\end{figure}
Figure~\ref{fig:architecture} shows an architecture where our approach fits. Starting from a process modeler, designers can create new process definitions or modify existing ones. The process definition is the input to the \texttt{Orchestration Rules Generator}. The generator instantiates predefined CQL templates for the modeling constructs of the respective process modeling language (Definition~\ref{def:abstract}). The generated CQL rules (statements) are bound to the nodes of the model by their IDs and the respective process model ID. This binding avoids the confusion of the same construct type among multiple models deployed for execution. The orchestration service (engine) \emph{orchestrates} the execution of the different process instances by watching the generated events over an event cloud, e.g., a message broker, matching these events with triggers for the deployed CQL rules. The engine updates the state table and generates more events that can be consumed internally or thrown to the event cloud so that external agents interested in the event can consume it, do their logic, and then generate other events. The architecture lends itself natively to a microservices architecture, which provides the most decoupling between the components. Moreover, changes in the process model can be easily deployed to the engine with full control of the logic of migrating running instances (more on this in Section~\ref{sec:flexibility}).
\begin{table}[t]

    \centering
    \includegraphics[width=\linewidth]{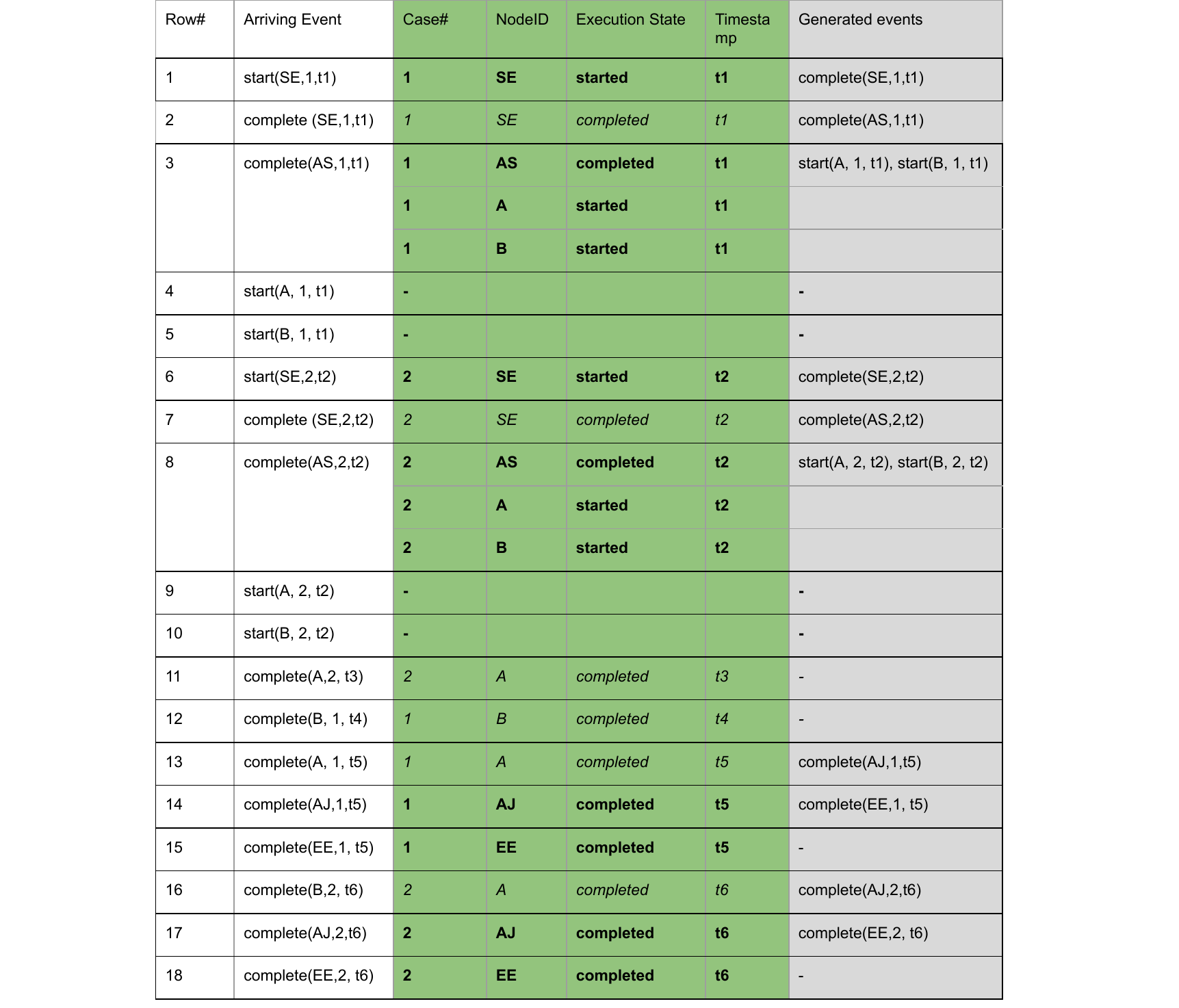}
    \caption{State table upserts based on the events generated in Fig.~\ref{fig:execution}. The left-most and right-most columns are for the arriving and the generated events respectively. The middle four columns represent the state table as per Definition~\ref{def:event:window}. Inserted tuples are in bold. Updated tuples are in italics. Events that do not upsert the table are marked with $-$.}
    \label{table:upserts}
\end{table}

As an example of executing a BPMN model, \figurename~\ref{fig:execution} shows a timeline of an example execution of the sample process model in \figurename~\ref{fig:architecture}, which consists of an interleaving execution of activities A and B\footnote{An imperative model is used for illustration purposes only.}, whereas~\figurename~\ref{table:upserts} shows the upserts to the state table and the generated events in response to arriving events. An agent triggers a new instance by sending an event $start(SE,1, t1)$. $SE$ is the node identifier that was communicated to the agent by sharing the process model or relevant parts thereof~\cite{DBLP:conf/bpm/GrunertSRPG22}. The number $1$ is the identifier assigned to the new process instance. The event is generated at time $t1$. The event is consumed by a CQL rule deployed to the engine (cf. Sec.~\ref{sec:rule:generation},~\ref{sec:declarative:rule:generation}). Such a CQL rule results in inserting a new tuple into the state table. Moreover, the engine generates four events $complete(SE,1, t1)$ and $complete(AS,1, t1)$, which indicate the completion of the start event and the \texttt{AND-Split}, respectively,  $starte(A,1, t1)$, and $starte(B,1, t2)$. \figurename~\ref{table:upserts} shows that these result in inserts and generation of other events, rows $1-3$.  The former two events are internal events consumed by other CQL rules. $complete(SE,1,t1)$ is generated in response to $start(SE,1,t1)$, row $1$, and $complete(AS,1,t1)$ is generated in response to $complete(SE,1,t1)$, row $2$. $start(A,1,t1)$ and $start(B,1,t2)$ are generated in response to $complete(AS,1,t1)$, row $3$. All these events are generated instantly and inherit the same timestamp as the $start(SE,1,t1)$. We notice that events $start(A,1,t1)$ and $start(B,1,t1)$ do not change the state table. This means that the engine, via CQL rules, does not respond to these events. Rather, as shown in ~\figurename~\ref{fig:execution} an external agent, e.g., performer work list, is listening to such events and can respond to them. At time $t2$, another process instance is created following the same sequence. At $t3$, task $A$ of the process instance $2$ is completed. This engine response updates the state table, row $11$ in ~\figurename~\ref{table:upserts}. Referring to the model, it is not possible to trigger the synchronization of the $AJ$ \texttt{AND-join} as the completion event of task $B$ for process instance $2$ is not observed yet. At $t6$, $comlpete(B, 2, t6)$ is observed, and the engine can update the state and trigger the completion of the \texttt{AND-join} $AJ$, row $16$, and the \texttt{End Event} $EE$ for process instance $2$, row $18$. For process instance $1$, tasks $B$ and $A$ complete at times $t4$ and $t5$, respectively, when the whole instance is completed.


Although the example above is for BPMN as a procedural language, our generic approach covers declarative and hybrid process enactment. Detailed examples, as proof of the use of streams and tables for process enactment, are given in the following two subsections for BPMN (Section~\ref{sec:rule:generation}) and DCR graphs (Section~\ref{sec:declarative:rule:generation}) as two prominent procedural and declarative process modeling languages respectively. The sections cover the mapping of the execution semantics of the language constructs into CQL statements and the necessary state tables to track cases' progress. Hybrid process modeling and enactment are discussed in Section~\ref{sec:hybrid:enactment}.

\subsection{Rule Generation for Imperative Process Models: BPMN}
\label{sec:rule:generation}

\begin{figure}
    \centering
    \includegraphics[width=1\linewidth]{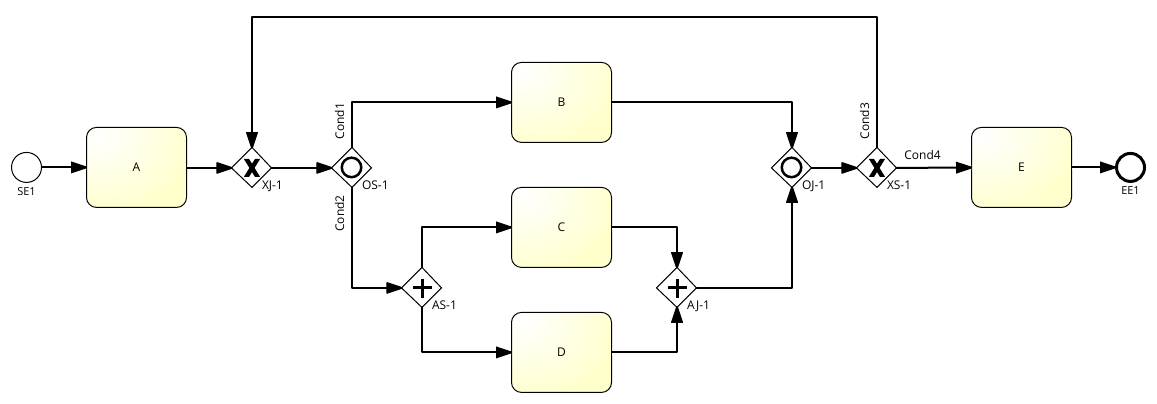}
    \caption{Example process model}
    \label{fig:running:example}
\end{figure}

We assume that process modeling guidelines are followed~\cite{MENDLING2010127}, especially using a single control flow edge for non-gateway nodes in the process model. As discussed in Section~\ref{sec:stream:table:duality}, the streams and tables represent a duality, where one can be reconstructed from the other. We capitalize on this fact to enact business processes. In our context, one stream is the \emph{raw} event stream, Definition~\ref{def:raw:event}, which manifests the change in the state of a process instance caused by executing a control flow node in the respective process model. The state table contains a snapshot of the evolution of cases. CQL provides the necessary language to define the schema of streams, windows, and tables. Moreover, CQL provides the language that allows responding to these events, which might result in updates to tables and/or new events generation on the stream.

Listing~\ref{sql:case:variables} defines one state table that will hold a snapshot of case variables. That is, the current value for each variable (data object) is defined in the process model. This table will be updated in response to \emph{completion} events received as the case (process instance) evolves.

\begin{lstlisting}[language=SQL,label={sql:case:variables}, caption={Case variables table}]
Create Table Case_Variables (pmID int primary key, caseID int primary key, variables Map);
\end{lstlisting}
Listing~\ref{sql:raw:event} defines the schema for the raw events that will be exchanged on the event stream. The schema follows Definition~\ref{def:event:stream}. The schema implicitly refers to the case variables via the \texttt{payLoad Map} column. The columns, \texttt{pmID, caseID} are used to correlate events and control which CQL rules to trigger, more details will be discussed shortly. \texttt{nodeID} and \texttt{state} are used to filter and lookup the state table.

\begin{lstlisting}[language=SQL, label={sql:raw:event}, caption={Raw execution event stream}]
Create Schema Process_Event
(pmID int, caseID int, nodeID string, state string, payLoad Map, ts Timestamp);
\end{lstlisting}


Listing~\ref{sql:execution:history} defines the dual table of the event stream. The table is updated in response to every received raw stream.
\begin{lstlisting}[language=SQL, label={sql:execution:history}, caption={Execution state table}]
Create Table Execution_State (pmID int primary key, caseID int primary key, nodeID string primary key, state string, timestamp long);
-- update the table in response of arriving events on the stream
On ProcessEvent as pe
merge Execution_State as es where es.pmID = pe.pmID and es.caseID = pe.caseID and es.nodeID = pe.nodeID
    when matched then
        update set es.state = pe.state, es.timestamp = pe.timestamp
    when not matched then
        insert select pe.pmID, pe.caseID, pe.nodeID,  pe.state, pe.timestamp;
\end{lstlisting}

In the following subsections, we capture the execution semantics of the different BPMN flow nodes (Definition~\ref{def:bpmn}) using CQL templates. Parameters to the templates are included within angle brackets. These parameters are substituted by their actual values based on the input process model.

    

    

\subsubsection{Mapping Events}
Start events trigger the creation of new process instances. The event's creation on the stream is assumed to be generated by an external agent, see \figurename~\ref{fig:execution}. The event is instantaneous. Therefore, once the engine detects a start from an agent referring to the respective start event node ID, it throws a \texttt{complete} event on the stream to execute the next nodes in the process. Any data payload in the start event is used to instantiate the case variables for the respective process instance by adding a new tuple to the \texttt{Case\_Variables} table. 

Formally, for a start event $se \in \mathcal{E} \wedge type(se) = start$, Definition~\ref{def:bpmn}, when a raw execution event  $(model, case, \mathcal{F}_e^{id}(se),started,payload,executor,ts)$ is observed, a new tuple is added to the state table, $\mathcal{T} = \mathcal{T} \cup \{(model, case, \mathcal{F}_e^{id}(se), started, ts)\}$ and a new event $(model, case, \mathcal{F}_e^{id}(se), completed, payload, executor, ts)$ is generated on the stream with all properties of the observed event except that the execution state will be \emph{completed}. 

List.~\ref{sql:start:event} shows the CQL rules that operationalize the response semantics to a start event. Such CQL rule is used to give the effect shown in row $1$ in Table~\ref{table:upserts}.

\begin{lstlisting}[language=SQL, label={sql:start:event}, caption={Handling a start event}]
Insert into Process_Event(pmID, caseID, nodeID,  "completed", payLoad, ts)
Select pmID, caseID, <StartEvent>, "completed", payLoad, ts 
from Process_Event(nodeID=<StartEvent>, state="started");

Insert into Case_Variables (pmID, caseID, variables)
Select st.pmID, st.caseID, st.payLoad from Process_Event(nodeID=<StartEvent>, state="started") as st;
\end{lstlisting}

Intermediate events can be handled as we handle tasks (the next section). List.~\ref{sql:end:event} handles end events. We focus on the control flow aspects as the engine waits for events generated by preceding nodes, typically one~\cite{MENDLING2010127}, to generate the end event on the stream. 

The generated execution event, the one to be inserted in the event stream at line $1$ will inherit the same process model ID and case ID as the received execution event of any of the previous nodes \texttt{pred} in the process model (Line 6). The value of the event's \texttt{state} will depend on the state of the \texttt{pred} event. If \texttt{pred} was \emph{skipped}, the state will be \emph{skipped} as well. On the other hand, if \texttt{pred} was \emph{completed}, the generated event will be \emph{completed} only if its incoming edge condition evaluates to \emph{true} using the case variables. This logic is shown in lines 3-5. This part of the CQL statement applies to all non-start nodes in a BPMN model. However, the logic of deciding to skip varies based on the execution semantics of the node. For instance, gateways have different logic for choosing their execution state as will be shown later.

Formally, for an end event $ee\in \mathcal{E} \wedge type(ee) = end $, an execution event $(model, case,\\ ee, completed, payload, executor, ts)$ will be observed on the event stream whenever an event $(model, case, prev, completed, payload, \_, ts)$ is observed on the stream where $(prev,ee) \in \mathcal{S}_f \wedge evaluate(exp((prev,ee)),payload)=true$. Otherwise, an execution event $(model, case, ee, skipped, payload, executor, ts)$ shall be observed. Similarly, if an event $(model, case, prev, skipped, payload, executor, ts)$ an event $(model, case, ee, skipped,\\ payload, executor, ts)$ shall be observed on the stream. List.~\ref{sql:end:event} gives the CQL rules that realize these semantics. Moreover, the statement clears the state table for the respective case only when an end event completion is observed.

\begin{lstlisting}[language=SQL, label={sql:end:event}, caption={Handling an end event}]
insert into Process_Event(pmID, caseID, nodeID,  state, payLoad, timestamp)
select pred.pmID, pred.caseID, <EndEvent>, 
case when pred.state= "completed" and evaluate(CV.variables, <Condition>) = true then "completed" else "skipped" end, CV.variables, pred.timestamp
From ProcessEvent as pred join Case_Variables as CV on pred.pmID = CV.pmID and pred.caseID = CV.caseID
where pred.state in ("completed", "skipped") and pred.nodeID in ([List of previous nodes of <EndEvent>]);

On Process_Event(nodeID=<EndEvent, state="completed") as end
Delete From Execution_State as H where H.pmID = end.pmID and H.caseID = end.caseID and H.nodeID <> end.nodeID and H.state ="completed";
\end{lstlisting}

We clear the respective state from the execution history as it is no longer needed and make room for future instances, lines $8$ and $9$. This is a separate CQL statement that will execute in response to the generation of \emph{completed} end event only. Clearing the ending case events from the memory does not necessarily mean losing the execution footprint. A logging agent (Figure~\ref{fig:architecture}) can listen to the generated events on the $Process\_Event$ stream and save a copy of these events to a log.

\subsubsection{Mapping Tasks}

The stream processing engine creates the raw execution event indicating the start of a task once the input control flow has been satisfied, rows $4,5,9,~and~10$ in Table~\ref{table:upserts} for example. However, this happens in case the preceding nodes have been completed. Otherwise, the engine generates a skip event. The latter case happens when, for instance, a conditional branch of the process is not satisfied. Referring to the example process in \figurename~\ref{fig:running:example}, if \texttt{Cond1} is not satisfied, the engine will generate an event $skip(B,\_,t_x)$. Otherwise, $start(B,\_,t_x)$ will be generated. Therefore, to allow the task to be executed, the preceding control flow node must generate a \texttt{completed} event, and the data condition must be satisfied. 

The upper CQL statement in Listing~\ref{sql:task} captures the execution semantics of tasks as described above. The lower statement is triggered upon the receipt of a \emph{completed} task execution event to update case variables from the event's payload. 

\begin{lstlisting}[language=SQL, label={sql:task}, caption={Handling tasks}]
Insert into Process_Event(pmID, caseID, nodeID, state, payLoad, ts)
Select pred.pmID, pred.caseID, <Task>, Case When pred.state="completed" and evaluate(CV.variables, <Condition>) = true then "started" else "skipped" end, CV.variables, pred.ts from Process_Event as pred join Case_Variables as CV on pred.pmID = CV.pmID and pred.caseID = CV.caseID where pred.state in ("completed", "skipped") and pred.nodeID = <PreviousNodeID>;

On ProcessEvent(nodeID=<Task>, state="completed") as event
Update Case_Variables as CV set variables(<Output>) = event.payLoad(<Output>) where CV.pmID = event.pmID and CV.caseID = event.caseID;
\end{lstlisting}

Receiving intermediate events needs both a control flow event and a message to arrive to be triggered. Without loss of generality, we can reuse the same logic as we handle tasks, where we can also extend the \texttt{where} clause to ensure that the respective message has arrived. This will require another table to track the messages. Throwing message events use the same logic as tasks, generating control flow events and messages. We can observe that the first CQL statements in List.~\ref{sql:end:event} and List.~\ref{sql:task} are identical as they two flow nodes have similar execution semantics. However, we kept both listings for the other statements that clear the state table (List.~\ref{sql:end:event}) and update the case variables (List.~\ref{sql:task}).

\subsubsection{Mapping Gateways}

In general, splitting gateways are handled the same way tasks are handled. They pass the same lifecycle state to their succeeding nodes. Note that in Listing~\ref{sql:task}, the condition of the input sequence flow to the task, if any, is evaluated, and its respective $start$ event is generated only if the evaluation succeeds, otherwise, the task is skipped. This implicitly handles cases of \texttt{XOR} and \texttt{OR} splits where sequence flow conditions are relevant.

Join gateways are handled following their type, i.e., an \texttt{AND}, \texttt{OR}, or an \texttt{XOR} gateway. For \texttt{AND} joins, the arrival of a control flow event of a preceding node will trigger the \texttt{AND} join node only if all preceding nodes' events have been received and have the same execution state, i.e. \texttt{completed} or \texttt{skipped}, as the most recently received event. Formally, for an \texttt{AND-join} flow node $j\in \mathcal{G} \wedge type(j) = AND$, a \texttt{completed} event $(model, case,j, completed, payload, executor, ts)$ will be observed on the execution stream if $\forall n : (n,j) \in \mathcal{S}_f,~\exists (model, case, \mathcal{F}^{id} _e (n), completed, ts_n) \in \mathcal{T}$, where $ts = Max(\{ts_n\})$. Similarly, a \texttt{skipped} event $(model, case,j, skipped, payload, executor, ts)$ will be observed on the execution stream if $\forall n : (n,j) \in \mathcal{S}_f,~\exists (model, case, \mathcal{F}^{id}_e (n),\\ skipped, ts_n) \in \mathcal{T}$, where $ts = Max(\{ts_n\})$.

List.~\ref{sql:and:join} shows the CQL statements that respond to the arrival of \texttt{AND-join} predecessors' events and trigger the respective \texttt{AND-join} execution state according to the logic above. The \emph{exists} sub-query from line $4$ is to be repeated for each preceding node to the AND-join. The logic checks for the existence of the most recent event for each preceding node whose state matches the triggering event state. Moreover, there has to be \emph{NO} AND-Join event, line $6$, already stored in the state table with a timestamp equal to or greater than any preceding nodes. This is particularly relevant in case the AND block is nested within a loop. Referring to the running example in \figurename~\ref{fig:running:example}, receiving the first $completed(C,x, t_n)$ event will not trigger the gateway for case $x$. Rather, the event will be kept in the state table. Once the event $completed(D,x,t_m)$ is received, the CQL statement in List.~\ref{sql:and:join} is matched and the new event for the completion of the AND join gateway is inserted into the $Process\_Event$ stream, allowing the process instance to proceed. We assume that once the task has been started, it has to be completed. That is, executors of the task instance cannot deliberately skip it. Skipping in the context of this paper is a decision made by the execution engine and is taken only in case the data condition for some sequence flow is not satisfied. Therefore the respective node is skipped and possibly all succeeding nodes. If a user can skip an in-progress task, another skipped event can be handled. 

Receiving a mix of \texttt{completed} and \texttt{skipped} execution events for the predecessors of an \texttt{AND-join} node indicates a modeling error, i.e. a deadlock in this case, as some execution branches leading to the \texttt{AND-join} can be skipped. Checking for modeling errors is beyond the scope of this work.

\begin{lstlisting}[language=SQL, label={sql:and:join}, caption={Handling AND joins}]
insert into ProcessEvent(pmID, caseID, nodeID,  state, payLoad, timestamp)
select pred.pmID, pred.caseID, <ANDJoin>, case pred.state when "completed" then "completed" else "skipped" end,pred.payLoad, pred.timestamp
from ProcessEvent(nodeID in (<List of previous nodes>)) as pred where  
  exists (select 1 from Execution_State as H where H.nodeID = <PreviousNode> and H.caseID = pred.caseID
    and H.state = pred.state and H.pmID = pred.pmID and H.timestamp <= pred.timestamp);
\end{lstlisting}

Handling \texttt{OR} joins is slightly different. The gateway is triggered with the last arriving execution event for all its predecessor nodes. However, it is sufficient to have one \texttt{completed} predecessor to generate a \texttt{completed} execution event for the \texttt{OR-join}. Otherwise, when all predecessors are skipped, the \texttt{OR-join}'s execution will be \texttt{skipped}. Formally, for an \texttt{OR-join} flow node $j\in \mathcal{G} \wedge type(j)=OR$, a \texttt{completed} event $(model, case,j,\\ completed, payload, executor, ts)$ will be observed on the execution stream if $\forall n : (n,j) \in \mathcal{S}_f, \exists (model, case, \mathcal{F}^{id} _e (n), completed, ts_n) \in \mathcal{T}~\vee~\exists (model, case, \mathcal{F}^{id} _e (n), skipped,\\ ts_n) \in \mathcal{T}$, where $ts = Max(\{ts_n\})$. Similarly, a \texttt{skipped} event $(model, case,j, skipped,\\ payload, executor, ts)$ will be observed on the execution stream if $~\forall n : (n,j) \in \mathcal{S}_f |\\ \exists (model, case, \mathcal{F}^{id}_e (n),\\ skipped, ts_n) \in \mathcal{T}$, where $ts = Max(\{ts_n\})$. 

The respective CQL statements are shown in List.~\ref{sql:or:join}. Comparing to List.~\ref{sql:and:join}, the difference is at Line $5$. In the \texttt{AND-join} case, we expect predecessors to be either all \emph{completed} or \emph{skipped}, whereas we relax this condition for the \texttt{OR-join}. 

\begin{lstlisting}[language=SQL, label={sql:or:join}, caption={Handling OR joins}]
insert into ProcessEvent(pmID, caseID, nodeID,  state, payLoad, timestamp)
select pred.pmID, pred.caseID, <ORJoin>, case pred.state when "completed" then "completed" else "skipped" end,pred.payLoad, pred.timestamp
from ProcessEvent(nodeID in (<List of previous nodes>)) as pred where  
  exists (select 1 from Execution_History as H where H.nodeID = <PreviousNode> and H.caseID = pred.caseID     and H.state in ("skipped", "completed") and H.pmID = pred.pmID and H.timestamp <= pred.timestamp );
\end{lstlisting}

XOR joins are triggered for each arriving event from a preceding node copying the same state as either completed or skipped. The model must ensure that at most one branch is executed. Otherwise, several events will be generated on the stream. 

\begin{lstlisting}[language=SQL, label={sql:xor:join:block}, caption={Handling XOR joins}]
Insert into Process_Event(pmID, caseID, nodeID, state, payLoad, ts)
Select pred.pmID, pred.caseID, <XORJoin>, case pred.state when "completed" then "completed" else "skipped" end, null, pred.ts
  from Process_Event (nodeID in ([list of previous nodes of <XORJoin>])) as pred;
\end{lstlisting}

\subsubsection{Loops}
\label{sec:loops}


We consider XOR-joins as the entry points for loops. We distinguish two sets of inputs for an XOR-join, loop-less, and looping. For our example (\figurename~\ref{fig:running:example}), \texttt{XJ-1} is an entry to a loop. Activity \texttt{A} is a loop-less entry. It resembles the first triggering for the loop control flow. The looping entry for \texttt{XJ-1} comes from \texttt{XS-1}. The CQL template to handle the loop-less and the looping cases is based on the template in List.~\ref{sql:xor:join:block}. The difference in the case of the loop-less entries is that we restrict the \texttt{nodeID} to those that do not repeat with the loop, for example, activity \texttt{A} for the \texttt{XJ-1} XOR-join. For the looping part, we restrict the \texttt{nodeID} for the node IDs of the nodes with an incoming edge to the XOR-join, aside from the loop-less nodes. 

Unstructured loops, a.k.a irreducible loops, are handled similarly. In this case, there are one or more entries and exits to the loop.

\begin{figure}[t]
    \centering
    \includegraphics[width=\linewidth]{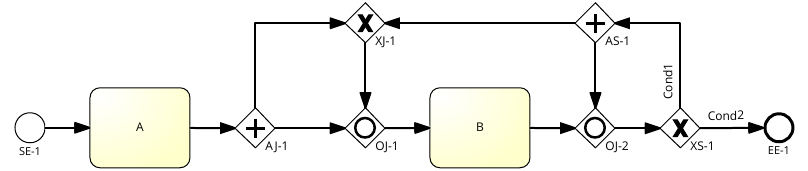}
    \caption{A process with irreducible loop}
    \label{fig:irreducible:loop}
\end{figure}

There could be cases where an OR-join is part of an unstructured loop. In this case, handling OR-joins following List.~\ref{sql:or:join} will cause a deadlock because not all incoming branches to the OR-join will have their state as either \texttt{completed} or \texttt{skipped}. ~\figurename~\ref{fig:irreducible:loop} shows a process model with an irreducible loop. In this process, \texttt{OJ-1} and \texttt{OJ-2} will not have all input branches either completed or skipped in all cases. 

If there is an input to an OR-join that is coming from a node that is a loop-less entry when the node is activated, we should not wait for other entries from the looping entries. Otherwise, we should wait for a \texttt{complete} or \texttt{skipped} state of all entries. For \texttt{OJ-1}, both \texttt{AJ-1} and \texttt{XJ-1} can be completed or skipped at the loop-less entry. Therefore, we should wait for tokens from both \texttt{AJ-1} and \texttt{XJ-1}. 
For future activations, we should wait for input only from nodes that are part of the loop. For \texttt{OJ-1}, this means input only from \texttt{XJ-1}. For \texttt{OJ-2}, for the first execution, i.e. loop-less entry, only  activity \texttt{B} is waited for. For further execution iterations, \texttt{OJ-2} waits for both \texttt{B} and \texttt{AS-1} are members of the loop structure. Therefore, \texttt{OJ-2} should wait for input from both.

The identification of input nodes that are loop-less or loop-entry is done at CQL statements generation time, i.e., based on the input process model structure. To identify these nodes, we employ a set of BPMN-Q~\cite{DBLP:journals/cii/AwadS12} queries that are tailored for this purpose. The interested reader is referred to our GitHub~\footnote{\url{https://github.com/AhmedAwad/Flexible-Process-Enactment-CEP}} repository for the details of these queries.


\subsection{Rule Generation for Declarative Process Models: DCR Graphs}
\label{sec:declarative:rule:generation}

This section uses DCR graphs (Definition~\ref{def:DCRLTS}) as a representative declarative business process modelling language. We show how CQL and tables can be used to orchestrate (execute) DCR graphs' processes. Following Definition~\ref{def:DCRLTS}, each event state can be described by three Boolean variables to indicate whether the event has \emph{happened} ($h$), is \emph{restless} ($r$), i.e. is still pending execution, or is \emph{included} ($i$). Therefore, we adapt the state table (Definition~\ref{def:event:window}) to accommodate the three boolean variables used to capture the state. Note that overall, the notion of the state table is still the same as the execution state of an activity is an abstract concept that materializes in the context of the process modeling and the execution language. The refined state table is introduced formally in Definition

\begin{definition}[DCR (State) Table]\label{def:event:window:dcr} Let $S$ be an event stream and let $\mathrm{T}$ be the set of timestamps, a state table $\mathcal{R}$ is relational table , $\mathcal{R} \subseteq \mathcal{M} \times \mathcal{C} \times \mathcal{N}\times \mathcal{H} \times \mathcal{I} \times \mathrm{R} \times \mathrm{T}$ whose primary key $PK \subseteq \mathcal{M} \times \mathcal{C} \times \mathcal{N}$, where $\mathcal{M}$ is the universe of process models, $\mathcal{C}$ is the universe of case identifiers, and $\mathcal{N}$ is the universe of nodes identifiers within process models, $\mathcal{I}$ indicates whether the DCR event happened, $\mathcal{I}$ indicates whether the DCR event is included, and $\mathrm{R}$ indicates whether the event is required. $\mathcal{H} = \mathcal{I} = \mathrm{R} = \{0,1\}$. The arrival of a new event $e \in \mathcal{S}$ may result in an upsert operation into $\mathcal{R}$.
\end{definition}

a state table can be defined as shown in Listing~\ref{sql:DCR:state:table}, the $EventState$ table indeed materializes the marking of a DCR graph instance. The execution events can use the same schema as in Listing~\ref{sql:raw:event}.

\begin{lstlisting}[language=SQL,label={sql:DCR:state:table}, caption={DCR Event State table}]
create table EventState (ProcessModelID long primary key, caseID long primary key, eventID string primary key, happened boolean, included boolean, restless boolean, ts Timestamp);
\end{lstlisting}

The $EventState$ is populated with records for each event in the DCR graph model. The initial marking of the model determines which events are included and pending or excluded, and this is not required (Listing~\ref{sql:DCR:initial:state}).
\begin{lstlisting}[language=SQL,label={sql:DCR:initial:state}, caption={Initialisation of the DCR Event State table}]
-- Initially Included Events
insert into EventState(ProcessModelID,caseID,eventID,happened,included,restless)
select <PM_ID>,<CASE_ID>,<INCLUDED_EVENT>,false,true,true;
-- Initially Excluded Events
insert into EventState(ProcessModelID,caseID,eventID,happened,included,restless)
select <PM_ID>,<CASE_ID>,<EXCLUDED_EVENT>,false,false,false;
\end{lstlisting}

Four relations in DCR graphs guide the execution of a DCR process: condition, response, include, and exclude (Definition~\ref{def:DCR:graphs}). Actions on event states define the new state that can be reached (Definition~\ref{def:DCRLTS}). The new state defines whether an event has occurred ($h$), is still pending ($r$), and is still included ($i$). The values assigned to these Boolean variables are determined by the semantics of the eligible relation from the current state and the action being executed. 

Formally, observing an execution event $e=(model,case,node,completed,payload,\\ executor, ts)$ on the stream, the respective tuple $t\in \mathcal{T}: t.eventID=e.node \wedge t.model = e.model \wedge t.case = e.case$ will be updated to $t^\prime$ where $t^\prime.model := t.model$, $t^\prime.case := t.case$, $t^\prime.node := t.node$, $t^\prime.happened=1$, $t^\prime.restless=0$ only if $t.included=1$. That is, the respective tuple in the event state table will be updated to indicate that the DRC event has happened and is no longer required only if it was included for execution and the respective execution events were observed with a \texttt{completed} state. 

If the execution event $e$ is involved in response relations, i.e., $e.node~\responseRel~target$, the target's tuple should be updated to indicate that it is required, only if it is included. That is, $r\in \mathcal{T}$, where $r.model = t.model \wedge r.case = t.case~\wedge~r.eventID=target$ will be updated to $r^\prime$ where $r^\prime.restless :=1$ only if $r.included = 1$. The execution of includes $\inclusionRel$ and exclude $\exclusionRel$ relations can be described in a way similar to the response relation where the $included$ flag is set accordingly.

If $e.node$ is conditioned, i.e., $src~\conditionRel e.node$, we can only update the respective tuple $t$ of the event $e$ if all $src$ events have their tuples either not included or both included and happened, where $t.included=1$. Formally, $t^\prime.happened :=1$, $t^\prime.restless=0$ only if $\forall src \in \Events : src \conditionRel e.node, \exists~st\in \mathcal{T} : st.model = t.model \wedge st.case = st.case~\wedge~st.eventID = src~\wedge (st.included=0 \vee (st.included=1 \wedge st.happened=1 )) \wedge t.included=1$. 

The following listings template CQL statements for the different relations. At least two updates take place on the $EventState$ table (Listing~\ref{sql:DCR:state:table}) upon the receipt of a raw execution event (Definition~\ref{def:raw:event}), the $EventState$ table is updated to reflect the execution of the received event and action. Other tuples in the $EventState$ are updated following the relations that are triggered by the received event.

\begin{lstlisting}[language=SQL,label={sql:DCR:relations}, caption={DCR Event State table updates}]
-- Update the executed activity/event to be happened (executed) and no longer required (restless=false)
on ProcessEvent(nodeID=<RawEvent.node>) as a
update EventState as ES set restless = false, happened=true
where ES.pmID = a.pmID and ES.caseID = a.caseID and ES.eventID=<RawEvent.node> and ES.included;

--Response relations triggered by the executed event response(source, target)
on ProcessEvent(nodeID=<source.node>) as a
update EventState as ES set restless = true,
where ES.pmID = a.pmID and ES.caseID = a.caseID and ES.eventID=<target.node> and ES.included;

-- Condition relations triggered by the executed event condition(conditioned, condition)
on ProcessEvent(nodeID=<conditioned>) as a
update EventState as ES set restless = false, happened=true,
where ES.pmID = a.pmID and ES.caseID = a.caseID and ES.eventID=<conditioned> and ES.included
    and exists (select 1 from EventState as ES2 where ES2.eventID = <condition> and ES2.caseID = ES.caseID
    and (not ES2.included or (ES.included and ES2.happened));

-- Includes relations triggered by the executed event includes(event, toInclude)
on ProcessEvent(nodeID=<event>) as a
update EventState as ES set included=true
where ES.pmID = a.pmID and ES.caseID = a.caseID and ES.eventID=<toInclude> 
and exists (select 1 from EventState as ES2 where ES2.pmID = a.pmID and ES2.caseID = a.caseID and ES2.eventID=a.nodeID and ES2.included=true);

-- Excludes relations triggered by the executed event includes(event, toExclude)
on ProcessEvent(nodeID=<event>) as a
update EventState as ES set included=false
where ES.pmID = a.pmID and ES.caseID = a.caseID and ES.eventID=<toExclude>
and exists (select 1 from EventState as ES2 where ES2.pmID = a.pmID and ES2.caseID = a.caseID and ES2.eventID=a.nodeID and ES2.included=true);

\end{lstlisting}

The \emph{response}, \emph{includes}, and \emph{excludes} relations in Listing~\ref{sql:DCR:relations} are straightforward. The \emph{condition} relation updates the state of the occurring \emph{conditioned} event to be executed and is no longer required only if all condition (preceding) events are either excluded or included and have already occurred. The nested query in the respective CQL statement will be repeated for all preceding condition events. 



\section{Flexibility: Enacting Hybrid Process Models}
\label{sec:hybrid:enactment}

In the previous section, we discussed how event streams, tables, and CQL are used to orchestrate BPMN models as a procedural modeling language and DCR graphs as a declarative language. In this section, we sketch and discuss how our approach can be used to orchestrate hybrid process models. 

A hybrid process model is intended to get the best of the two paradigms. Well-defined behavior can be captured through explicit control flow defining what is allowed, i.e. procedural. In contrast, highly-variant parts and less structured behavior can be captured employing declarative constraints. To this end, hybrid process models can broadly be classified as hierarchical hybrid models, e.g.,~\cite{DBLP:conf/dasfaa/AalstAHPS09,DBLP:conf/otm/SlaatsSMR16} or fully mixed models, e.g. ~\cite{mixedParadigms2013}. In literature, most of the hybrid approaches follow the hierarchical model.


Hierarchical hybrid models employ sub-processes as execution units, which can be modeled with procedural or declarative languages. BPMN \emph{ad-hoc sub-processes} partially support this. Despite formalized semantics, no engine can currently enact such models. The formalization approach emphasizes using observable behavior, which we connect to events. If sub-processes agree on event definitions, our approach enables events from procedural processes to be consumed by declarative ones and vice versa. We would like to investigate the correctness criteria and verification of hybrid models in future work.

A fully-mixed hybrid process model is achieved by projecting the procedural and declarative parts onto a richer modeling language. For instance, the work in~\cite{mixedParadigms2013}, projects Petri nets, DECLARE, and DCR constraints onto colored Petri nets (CPN). The resulting CPN model is guaranteed to satisfy the declarative and the procedural underlying models. As Tables in our approach can be defined to hold any data, the resulting CPN model can be translated into a set of rules following a similar approach to the one presented in Section~\ref{sec:rule:generation}.

Figure~\ref{fig:hybrid:case:management} depicts a hybrid process model where BPMN is the top-level process. The sub-process $P$ is a DCR graph. Following the execution semantics in~\cite{DBLP:conf/otm/SlaatsSMR16}, we discuss how our approach can be used to execute such a hybrid process.

Rules to orchestrate the procedural part will follow the templates discussed in 
Listings~\ref{sql:start:event} (start event),~\ref{sql:task}(tasks $A$ and $F$), and~\ref{sql:end:event} (end event) respectively. Similarly, the execution rules of the inner DCR process will be defined following the template in Listing~\ref{sql:DCR:relations}. The connection between the two executions is the activation of the sub-process $P$. The sub-process will be activated upon the receipt of the complete event of task $A$. This activation must instantiate the DCR instance by initializing its respective state. This is illustrated in Listing~\ref{sql:init:inner:DCR}. The initialization of the state table for the inner DCR process takes place in response to the occurrence of the $start$ event of the sub-process $P$.

\begin{lstlisting}[language=SQL,label={sql:init:inner:DCR}, caption={Inistantiation and termination of the inner DCR Graph process}]
on ProcessEvent(eventID=P,state="started") as a
insert into EventState(ProcessModelID,caseID,eventID,happened,included,restless)
select a.pmID, a.caseID, "B", false, true, true,
select a.pmID, a.caseID, "C", false, false, false,
select a.pmID, a.caseID, "D", false, true, true,
select a.pmID, a.caseID, "E", false, true, true;
\end{lstlisting}

The DCR process will respond to the occurrence of (external) events related to tasks, $B$, $C$, $D$, and/or $E$. Following~\cite{DBLP:conf/otm/SlaatsSMR16}, one or more actions, i.e., activities, need to signify the termination of the DCR process. In our example, we can assume that the occurrence of activity $D$ is such a terminating action. Note that, before observing the completion of $D$, activities, $B$, $C$, and $E$ can occur an arbitrary number of times, following a legitimate execution deduced from the declarative rules. Upon receipt of the execution event of $D$, the state table of the DCR process must be erased to prevent further execution of the declarative part of the process. Additionally, a $complete$ event of $P$ will be generated. Then, the BPMN process will proceed with activity $F$ until its end event.

\section{Flexibility: Handling Changes in the Model}
\label{sec:flexibility}

Changes in the process model can be due to many reasons, including improvements, compliance with regulations and internal policies, or due to feedback from process mining activities. Technically, changes in the model will reflect on the running cases instantiated from the affected model. Instances could be kept following the model \emph{before} change, or they might be migrated to the model \emph{after} change~\cite{DBLP:journals/ife/DadamR09,DBLP:conf/wecwis/Rinderle-MaR10}. For model changes occurring with running instances, it will be necessary to ensure consistency between pre- and post-states during the adaptation process~\cite{nahabedian2022assured}.

In our approach, the atomic deployment artifact is a CQL statement. Each statement is responsible for executing, i.e., consuming the input events (tokens) and producing the output event (token). Therefore, changes in the process model can divide the nodes in the process model into unchanged, added, removed, or modified. Only the rules of the three latter cases need to be updated on the stream processing engine. Rules are tracked by their IDs, inherited from the respective control flow node IDs. Therefore, it is possible to undeploy old rules and deploy new ones. However, more sophisticated migration situations might need to be supported. For instance, we might want to keep all running instances following the old model and only allow new instances to follow the new model. We argue that we can easily achieve that by adding filter conditions to the CQL rules that direct the process instances to the right rule.

Assuming a modification to the process model in \figurename~\ref{fig:running:example}, where a new parallel activity, \texttt{F}, is added within the AND-block and activity \texttt{E} is removed. In this scenario, the altered node is \texttt{AS-1}, gaining an additional outgoing edge, while the newly added node is the activity \texttt{F}, and the dropped node is \texttt{E}. Furthermore, the end event \texttt{EE1} is considered changed as it now follows \texttt{XS-1} instead of \texttt{E}. Changes to the output control flow of \texttt{AS-1} don't necessitate adjustments to the relevant rule, as the rule primarily concerns the input control flow. Deleting \texttt{E} may lead to the deletion of its respective rule if an immediate change is required for all running instances; otherwise, the rule will be redeployed following the template in List.~\ref{sql:task}. For \texttt{EE1}, the old rule is modified and redeployed to receive input from \texttt{E} for old instances. A new rule is also deployed to the same event to wait for events generated by \texttt{XS-1}. The filter condition will be the complement in that case, i.e., \texttt{pred.caseID > [ID]}. Deploying the new activity \texttt{F} involves creating a new rule, potentially with a filter on case identifiers based on the cases it needs to be enforced upon.

\section{Evaluation}
\subsection{Implementation}
\label{sec:implementation}

We show the feasibility of our approach via an implementation using the Java Programming language, and Esper\footnote{\url{https://www.espertech.com/esper/}} as a stream processing engine. In our implementation, we considered imperative BPMN models modeled in the Signavio process modeler, with Camunda libraries to read XML-based BPMN files and translate them into CQL statements. To identify loops and other complex structures, c.f. Section~\ref{sec:loops}, we used BPMN-Q~\cite{DBLP:journals/cii/AwadS12} queries to identify loop entry and exit nodes and other block types. The implementation of declarative process models included models DCR graphs developed in the DCR-js process modeller~\cite{tamo2023open}, generating XML files that were converted to CQL statements. More details on the implementation can be found in our GitHub repository~\footnote{\url{https://github.com/AhmedAwad/Flexible-Process-Enactment-CEP}}.
To address scalability, we leverage Esper's built-in context feature. Context is a means for multi-threading processing. A context is created for each case identifier. Internally, Esper shards the events by their caseID property, and depending on the available threads, the processing is forwarded to the available threads.
To evaluate conditional expressions in BPMN, i.e., control flow edge conditions, we generate Javascript functions that are interpretable by Esper. All functions have the defined variables as input parameters and the expression of \texttt{conditionExpression} is the function's body.
To control the growth of the execution history table, we purge events from the table whenever a process instance reaches an end event. There could be other policies to follow to purge events earlier. For example, the execution history can be time-based, assuming a maximum duration for any event to arrive. However, the wrong time window selection might result in deadlocks and failure to complete a case. However, our approach is flexible because the tables (windows) can be redefined, and events can be moved from the old to the new table.

\subsection{Case Study: A Case Management Business Process}\label{sec:processReq}

    In this section, we validate our approach by means mapping a process model from the literature into a set of CQL statements. We start by discussing the requirements for the process. Then, we develop a procedural BPMN process, a declarative DCR process, and a hybrid model combining imperative and declarative aspects,  and show how they can be mapped to CQL and their execution semantics. Moreover, we compare the execution of our BPMN and DCR processes to those deployed on Camunda and DCRGraphs, respectively, to show that we can reach similar execution sequences. The comparisons were on selected execution scenarios and are not meant to show full equivalence, this is a subject for future work.

    \begin{figure*}[t]
    \centering
    \includegraphics[width=.9\linewidth]{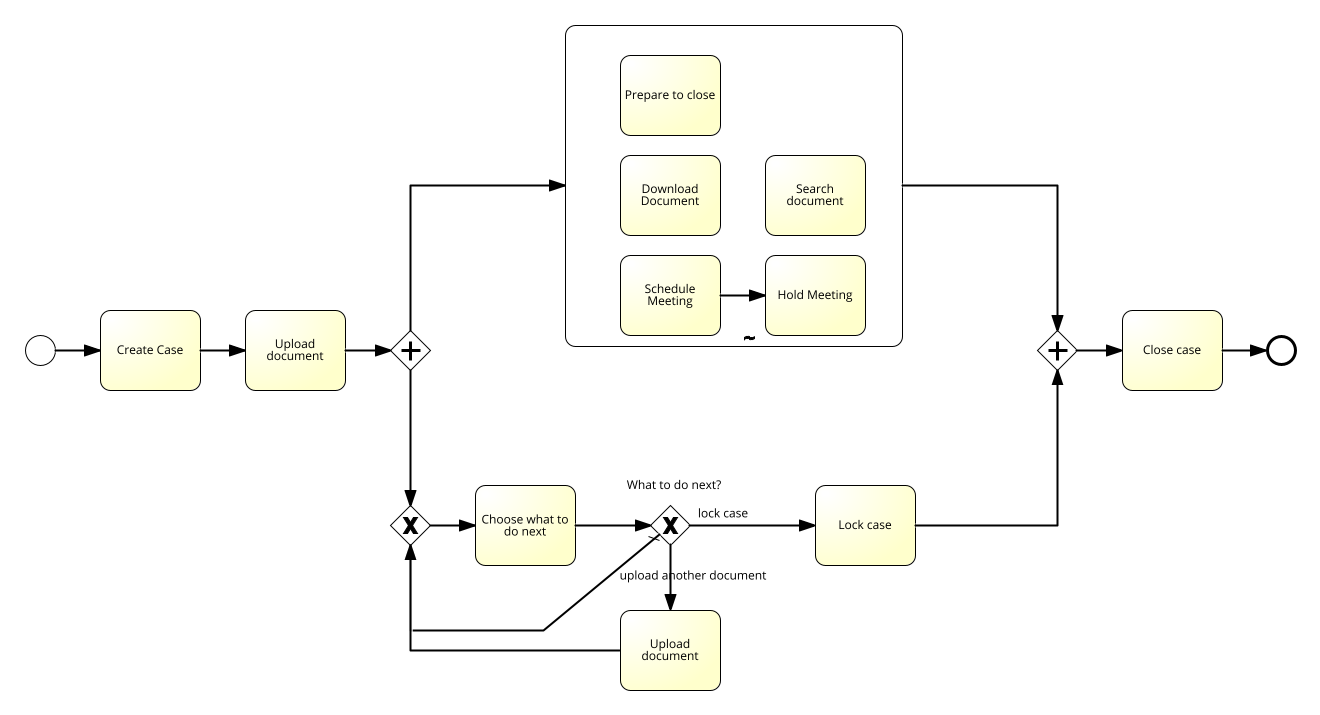}
    \caption{BPMN model of a case management process (reproduced from ~\cite{DBLP:journals/jodsn/Slaats20} Figure 6)}
    \label{fig:bpmn:case:management}
\end{figure*}

We use the process description from ~\cite{DBLP:journals/jodsn/Slaats20} of a  case management process with the following requirements: \begin{enumerate}
    \item Every case of the case management system is initially created and eventually closed,
    \item For a created case, an arbitrary number of documents can be uploaded,
  \item An uploaded document can be downloaded or searched,
  \item At any time, a case can be locked, 
  \item After locking a case, it is not possible to upload a document; still, uploaded documents can be downloaded and searched, 
  \item Furthermore, in every case, meetings can be held. To hold a meeting, it has to be (re) scheduled, 
  \item Meetings can be rescheduled arbitrarily often, however, it is not possible to schedule more than one meeting in advance. 
\end{enumerate}
The process has parts that can be best modeled following a procedural approach, i.e., the explicit start (creating a case) and end (closing a case), and another part that can be better captured by embracing a declarative approach, i.e., uploading, downloading documents, and scheduling and conducting meetings. In the following, we show how the process can be represented using BPMN and DCR graphs, and how it can benefit from using a hybrid approach. The objective is not to discuss the expressiveness of the modeling languages. Rather, we evaluate the flexibility of our approach for executing business processes.

\subsubsection{BPMN}

Figure~\ref{fig:bpmn:case:management} captures the process requirements outlined in Section \ref{sec:processReq} with the following observations:
\begin{itemize}
    \item The upload document task is duplicated. The first one is to force a document to be uploaded before any further actions can be taken on the case. The second copy of the task is to allow the optional behavior of uploading other documents,
    \item Two artificial tasks were introduced to make the process semantics clearer,
        \begin{itemize}
            \item The task ``Choose what to do next'' is introduced in a looping behavior to give the process performer the chance to either upload another document, lock the case, or do nothing. If she chooses to upload a document or do nothing, she will be offered to execute the ``Choose'' task again. If the case is locked, the control flow token is passed to the AND-join and waits for the termination of the other parallel branch.
            \item The task ``Prepare to close case'' is added to the ad-hoc sub-process to force a clear termination condition for the sub-process. Otherwise, one of the tasks among ``download document'', ``search document'', or ``schedule meeting'' can be executed.
        \end{itemize}
\end{itemize}

This model exposes additional behavior that is not in the original description. Both tasks ``Choose what to do next'' and ``Prepare to close case'' are not genuinely parts of the requirements and do not contribute a business value. Additionally, notice that the BPMN process does not fully capture the constraints in the process description in Section \ref{sec:processReq}. The description states that it is not possible to schedule more than one meeting in advance. If we look at the execution semantics of the BPMN Ad-hoc subprocess, we can execute the ``Schedule Meeting'' task several times before we conduct the meeting.

\begin{figure*}[t!]
    \centering
    \includegraphics[angle=-90,width=0.7\textwidth]{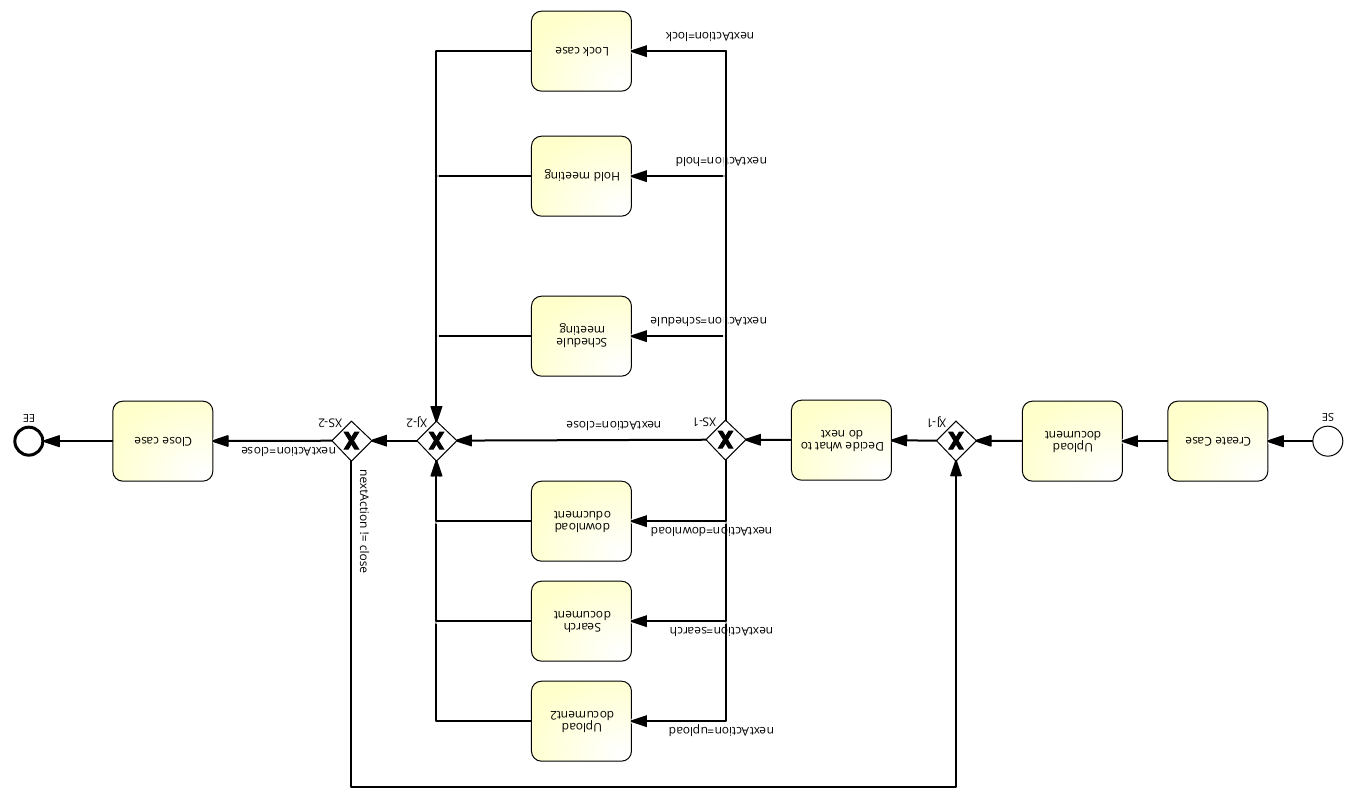}
    \caption{A BPMN model for case management - simplified}
    \label{fig:bpmn:case:management:simplified}
\end{figure*}

To stick to common BPMN elements that are truly procedural, we get rid of the Ad-hoc sub-process and remodel the process from Figure~\ref{fig:bpmn:case:management} as shown in Figure~\ref{fig:bpmn:case:management:simplified}. In the latter model, we get rid of the artificial task "Prepare to close case". Yet, we lose the parallelism of uploading documents, searching documents, planning, and holding meetings. The simplification allows modeling this process in other procedural process modeling languages and for comparison with other BPMN-compliant execution engines, e.g., Camunda~\footnote{Camunda does not support the execution of Ad-hoc sub-processes. Moreover, its XML parser does not recognize the respective XML tag.}.
The model in Figure~\ref{fig:bpmn:case:management:simplified} uses variables to control the execution flow. The logic of the "Decide what to do next" sets the value of the \textit{nextAction} variable. Moreover, the logic for forcing scheduling a meeting before holding it, blocking documents upload once the case is locked, and not scheduling more than one meeting are all decided within the "Decide what to do next" task.

\subsubsection{DCR} 
\begin{figure}[t]
    \centering
    \includegraphics[width=1\linewidth]{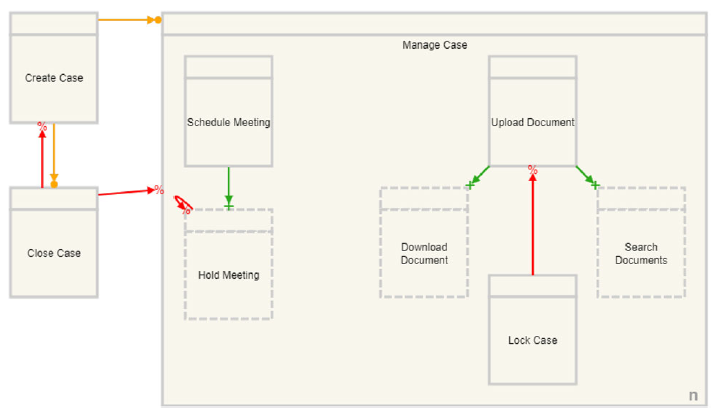}
    \caption{A DCR model for case management}
    \label{fig:dcr:case:management}
\end{figure}

We can illustrate corresponding modeling in a declarative process model using DCR graphs in Figure~\ref{fig:dcr:case:management}. The model is less complex, with no duplicate or artificial tasks. However, it is not easy to capture the starting and ending of the case. It takes a while to figure out that you can only execute ``Create Case'' at the beginning because it is set as a condition for all other events in the model. However, nothing prevents executing ``Create case'' several times for the same process instance (case). This can be remedied by adding an “exclude” relation to itself. The same happens for the ``Close case'' and ``Lock case''. Therefore, there are no clear termination conditions for such a model, as is the case for DCR graphs in general.

\subsubsection{The Hybrid Model} 
\begin{figure}[th!]
    \centering
    \includegraphics[width=1\linewidth]{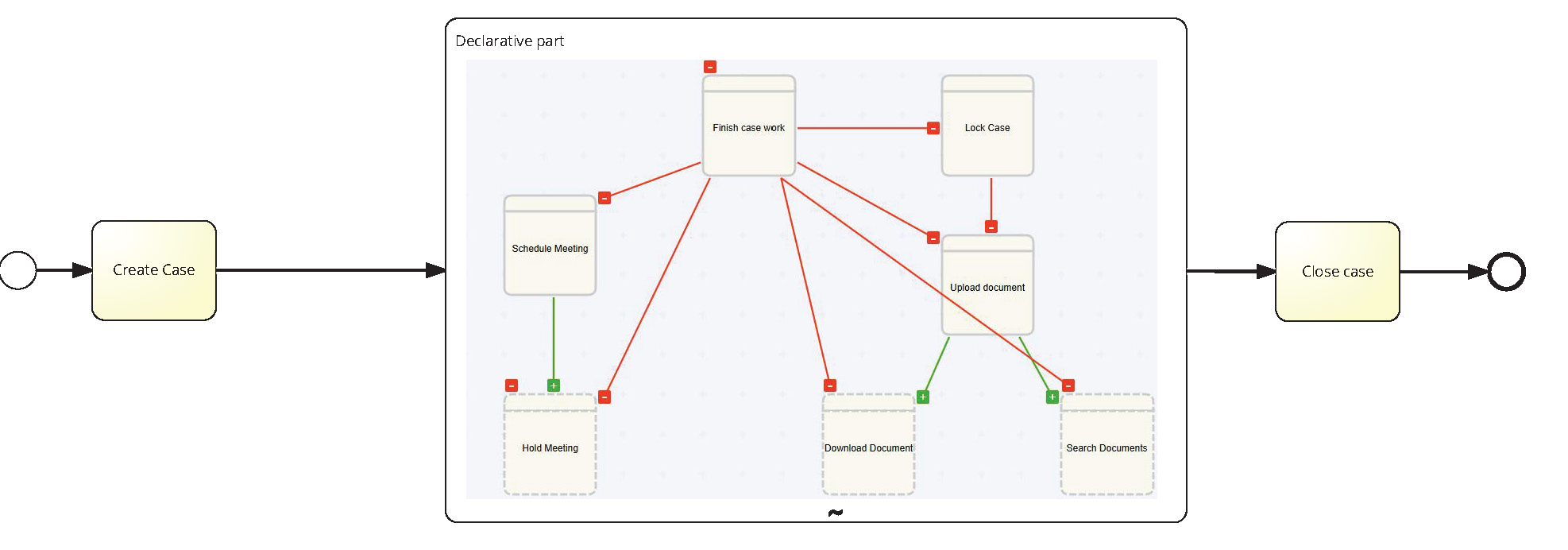}
    \caption{A hybrid model for case management}
    \label{fig:hybrid:case:management}
\end{figure}

Figure~\ref{fig:hybrid:case:management} assumes a nested approach for modeling hybrid process models. The model uses imperative processes at the top level, in this case, a BPMN process model, and DCR graphs for the declarative part. The two steps of creating and closing the case take place in the beginning and the end, respectively. The ad-hoc sub-process will host the declarative part. We can observe that overall, the new model is simpler and has fewer artificial tasks. We only introduce the “Finish case work” task to explicitly define termination conditions for the ad-hoc sub-process.

We have omitted the generated CQL statements due to space limitations. However, the full list of CQL commands can be found on our \href{https://github.com/AhmedAwad/Flexible-Process-Enactment-CEP/blob/4a5a3582643adeba4015a958891ead670fbf55b4/CaseStudy/CaseStudy.md}{GitHub repository}.

\subsection{Comparing with Camunda}

To further validate our approach, we have implemented the BPMN process in Figure~\ref{fig:bpmn:case:management:simplified} as a Camunda BPMN process. Task "Decide what to do next" was implemented as user task that is linked to a form where the user can choose the action to take next, whether to lock the case, and whether to have a further iteration. In our approach, we have implemented a simple command line tool to receive the user input for these decisions. Figure

\begin{figure}[htb!]
    \centering
    \includegraphics[width=\linewidth]{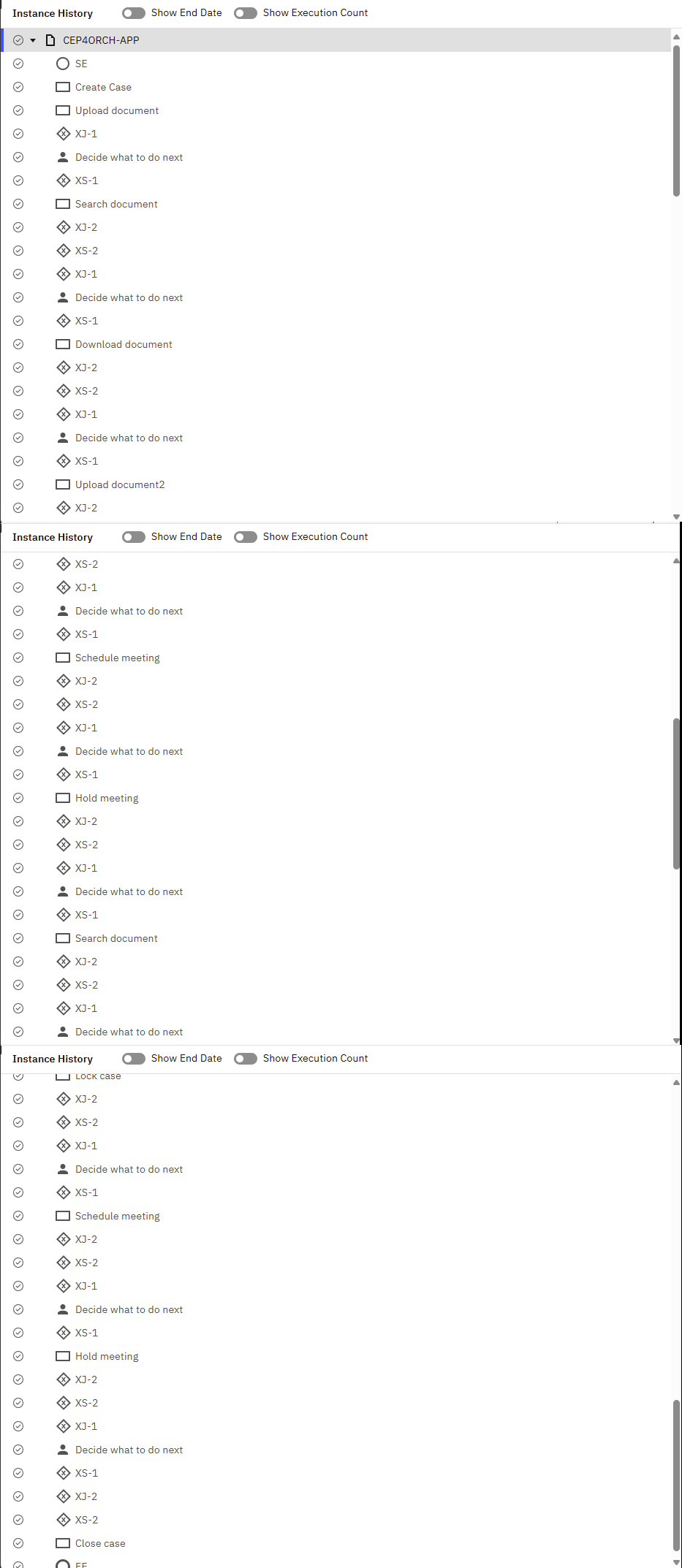}
    \caption{Camunda execution snapshot for one case of the BPMN process in Figure~\ref{fig:bpmn:case:management:simplified}}
    \label{fig:case:management:camunda}
\end{figure}

Listing~\ref{case:management:bpmn:log} shows the generated log from our execution. You can match the sequence of nodes executed to the one visualized in Figure~\ref{fig:case:management:camunda}. The freemium subscription level with Camunda does not allow us to extract the execution log. Each event in the listing follows the format of Definition~\ref{def:raw:event}~\footnote{We have omitted the timestamp as the execution order is conveyed in the order of the events in the listing. Moreover, we kept only the \emph{completed} events, to reduce the size of the listing and to match Camunda's output}. The payload of each event is shown as a dictionary of key-value pairs. For instance, at line $5$, the user has decided to search for a document as the \texttt{nextAction=search} pair indicates. 

\begin{lstlisting}[
  keywordstyle=\color{keywordcolor}\bfseries, % Custom keyword style
  morekeywords={completed, noiterate, false, close, upload, search, hold, true, schedule,lock,download }, % Custom keywords
  stringstyle=\color{stringcolor}, % String color
  commentstyle=\color{commentcolor}, % Comment color
  numbers=left, % Line numbers
  numberstyle=\tiny\color{gray}, % Line number style
  breaklines=true, % Break long lines
  frame=single, % Frame around the code block
  showstringspaces=false, % Don't show spaces in strings,
  label={case:management:bpmn:log},
  basicstyle=\footnotesize\sffamily,
  captionpos=b,
  caption={A sample log using streams and tables to execute one instance of the BPMN process in Figure~\ref{fig:bpmn:case:management:simplified}}
]
3,1,SE, completed, {caseLocked=false; nextAction=close;}
3,1,Create Case, completed, {caseLocked=false; nextAction=close;}
3,1,Upload document, completed, {caseLocked=false; nextAction=close;}
3,1,XJ-1, completed, {caseLocked=false; nextAction=close;}
3,1,Decide what to do next, completed, {caseLocked=false; nextAction=search;}
3,1,XS-1, completed, {caseLocked=false; nextAction=search;}
3,1,Search document, completed, {caseLocked=false; nextAction=search;}
3,1,XJ-2, completed, {caseLocked=false; nextAction=search;}
3,1,XS-2, completed, {caseLocked=false; nextAction=search;}
3,1,XJ-1, completed, {caseLocked=false; nextAction=search;}
3,1,Decide what to do next, completed, {caseLocked=false; nextAction=download;}
3,1,XS-1, completed, {caseLocked=false; nextAction=download;}
3,1,Download document, completed, {caseLocked=false; nextAction=download;}
3,1,XJ-2, completed, {caseLocked=false; nextAction=download;}
3,1,XS-2, completed, {caseLocked=false; nextAction=download;}
3,1,XJ-1, completed, {caseLocked=false; nextAction=download;}
3,1,Decide what to do next, completed, {caseLocked=false; nextAction=upload;}
3,1,XS-1, completed, {caseLocked=false; nextAction=upload;}
3,1,Upload document2, completed, {caseLocked=false; nextAction=upload;}
3,1,XJ-2, completed, {caseLocked=false; nextAction=upload;}
3,1,XS-2, completed, {caseLocked=false; nextAction=upload;}
3,1,XJ-1, completed, {caseLocked=false; nextAction=upload;}
3,1,Decide what to do next, completed, {caseLocked=false; nextAction=schedule;}
3,1,XS-1, completed, {caseLocked=false; nextAction=schedule;}
3,1,Schedule meeting, completed, {caseLocked=false; nextAction=schedule;}
3,1,XJ-2, completed, {caseLocked=false; nextAction=schedule;}
3,1,XS-2, completed, {caseLocked=false; nextAction=schedule;}
3,1,XJ-1, completed, {caseLocked=false; nextAction=schedule;}
3,1,Decide what to do next, completed, {caseLocked=false; nextAction=hold;}
3,1,XS-1, completed, {caseLocked=false; nextAction=hold;}
3,1,Hold meeting, completed, {caseLocked=false; nextAction=hold;}
3,1,XJ-2, completed, {caseLocked=false; nextAction=hold;}
3,1,XS-2, completed, {caseLocked=false; nextAction=hold;}
3,1,XJ-1, completed, {caseLocked=false; nextAction=hold;}
3,1,Decide what to do next, completed, {caseLocked=false; nextAction=search;}
3,1,XS-1, completed, {caseLocked=false; nextAction=search;}
3,1,Search document, completed, {caseLocked=false; nextAction=search;}
3,1,XJ-2, completed, {caseLocked=false; nextAction=search;}
3,1,XS-2, completed, {caseLocked=false; nextAction=search;}
3,1,XJ-1, completed, {caseLocked=false; nextAction=search;}
3,1,Decide what to do next, completed, {caseLocked=false; nextAction=lock;}
3,1,XS-1, completed, {caseLocked=false; nextAction=lock;}
3,1,Lock case, completed, {caseLocked=true; nextAction=lock;}
3,1,XJ-2, completed, {caseLocked=true; nextAction=lock;}
3,1,XS-2, completed, {caseLocked=true; nextAction=lock;}
3,1,XJ-1, completed, {caseLocked=true; nextAction=lock;}
3,1,Decide what to do next, completed, {caseLocked=true; nextAction=schedule;}
3,1,XS-1, completed, {caseLocked=true; nextAction=schedule;}
3,1,Schedule meeting, completed, {caseLocked=true; nextAction=schedule;}
3,1,XJ-2, completed, {caseLocked=true; nextAction=schedule;}
3,1,XS-2, completed, {caseLocked=true; nextAction=schedule;}
3,1,XJ-1, completed, {caseLocked=true; nextAction=schedule;}
3,1,Decide what to do next, completed, {caseLocked=true; nextAction=hold;}
3,1,XS-1, completed, {caseLocked=true; nextAction=hold;}
3,1,Hold meeting, completed, {caseLocked=true; nextAction=hold;}
3,1,XJ-2, completed, {caseLocked=true; nextAction=hold;}
3,1,XS-2, completed, {caseLocked=true; nextAction=hold;}
3,1,XJ-1, completed, {caseLocked=true; nextAction=hold;}
3,1,Decide what to do next, completed, {caseLocked=true; nextAction=close;}
3,1,XS-1, completed, {caseLocked=true; nextAction=close;}
3,1,XJ-2, completed, {caseLocked=true; nextAction=close;}
3,1,XS-2, completed, {caseLocked=true; nextAction=close;}
3,1,Close case, completed, {caseLocked=true; nextAction=close;}
3,1,EE, completed, {caseLocked=true; nextAction=close;}

\end{lstlisting}

\subsection{Comparing with DCRGraphs.net}
\begin{figure}[H]
    \centering
    \includegraphics[width=1\linewidth]{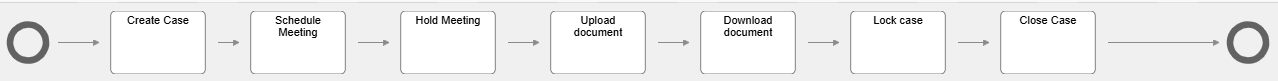}
    \caption{DCRGraphs.net run of the process}
    \label{fig:case:management:dcr:graph:net}
\end{figure}

We used the community edition. We have implemented the process in Figure~\ref{fig:dcr:case:management}. Figure~\ref{fig:case:management:dcr:graph:net} shows a run of the process which matches the log generated by our approach as shown in Listing~\ref{case:management:dcr:log}. After each execution, we list available tasks for execution. Checking for the enabled tasks for execution is achieved by querying the \texttt{EventState} table.

\begin{lstlisting}[language=, % Leave empty for generic text
  keywordstyle=\color{keywordcolor}\bfseries, % Custom keyword style
  morekeywords={Available, completed}, % Custom keywords
  stringstyle=\color{stringcolor}, % String color
  commentstyle=\color{commentcolor}, % Comment color
  numbers=left, % Line numbers
  basicstyle=\footnotesize\sffamily,
  numberstyle=\tiny\color{gray}, % Line number style
  breaklines=true, % Break long lines
  frame=single, % Frame around the code block
  showstringspaces=false, % Don't show spaces in strings,
  label={case:management:dcr:log},
  captionpos=b,
  caption={A sample log using streams and tables to execute one instance of the DCR process in Figure~\ref{fig:dcr:case:management}}
]
3,1,Create Case, completed, {None=none}, 1720343160040
	Available tasks are: Lock case, Close Case, Schedule Meeting, Upload document
3,1,Schedule Meeting, completed, {None=none}, 1720343165005
	Available tasks are: Lock case, Hold Meeting, Close Case, Upload document
3,1,Hold Meeting, completed, {None=none}, 1720343169731
	Available tasks are: Lock case, Close Case, Schedule Meeting, Upload document
3,1,Upload document, completed, {None=none}, 1720343181064
	Available tasks are: Lock case, Search documents, Close Case, Download document, Schedule Meeting, Upload document
3,1,Download document, completed, {None=none}, 1720343217725
	Available tasks are: Lock case, Search documents, Close Case, Download document, Schedule Meeting, Upload document
3,1,Lock case, completed, {None=none}, 1720343223139
	Available tasks are: Lock case, Search documents, Close Case, Download document, Schedule Meeting
3,1,Close Case, completed, {None=none}, 1720343235955
	Available tasks are: None
\end{lstlisting}

\section{Related Work}
\label{sec:related:work}

Defining and reasoning about business processes and workflow execution semantics is a well-established academic topic. The seminal work of van der Aalst~\cite{DBLP:conf/apn/Aalst97} proposed a special case of Petri nets (workflow nets) to represent business process models and to reason about the correctness of these models. This led to defining the notion of business processes \emph{soundness}~\cite{Groefsema2013ASO} utilizing the formal semantics of Petri nets. Petri nets have been used to formalize other process modeling languages like EPCs~\cite{DBLP:journals/cii/DongenJVA07} and BPMN~\cite{DBLP:journals/infsof/DijkmanDO08}, amongst other modeling languages. BPMN is currently the standard for business process modeling, so we focus our discussion to execution semantics and the engines supporting it. For declarative processes, we are only aware of the execution engine provided by DCR graphs~\cite{debois2016dcr}, which is not cross-compatible with BPMN models.

Providing execution semantics for advanced control flow constructs has been proposed for OR-joins~\cite{DBLP:conf/bpmn/GfellerVW11,omg2014bpmn}, unstructured loops~\cite{DBLP:conf/bpm/PrinzCH22}. Fahland and V\"{o}lzer~\cite{DBLP:journals/is/FahlandV18} proposed the use of so-called skip tokens and block guards to provide local semantics for OR-joins. The work brings the benefits of dead path elimination from BPEL to cyclic process models that contain OR-joins. The notion of skip token is very close to our approach in this paper. However, unlike~\cite{DBLP:journals/is/FahlandV18}, we use a declarative approach, SQL, and do not require an explicit extension of the modelling language with new constructs. Moreover, we provide support for case-level and task data as the aim is to build a process execution engine.

Caterpillar~\cite{DBLP:journals/spe/Lopez-PintadoGD19} is a blockchain-based BPMN execution engine. The control-flow routing is implemented as smart contracts that are generated by a BPMN-to-Solidity compiler. Caterpillar supports advanced constructs such as boundary events, different BPMN task types, and multiple-instance activities. However, unstructured loops and OR-joins are supported. Moreover, smart contracts are imperative commands, and support for changes in the model requires new deployments to the blockchain.

Following the notion of decentralization promised by blockchains, Grunert at al.~\cite{DBLP:conf/bpm/GrunertSRPG22} propose a decentralized architecture for the enactment of business processes by splitting and replicating the process model over several execution engines. The state of a process instance is migrated from one engine to another to guarantee execution continuation in case of the current host's failure. Process models are translated into Javascript code. Therefore, it still provides an imperative implementation for enacting processes. There is no discussion about how changes to the process model are handled.

Many other engines are available, some of them are proprietary software, and some are open source but, typically, they implement a single set of languages with these implementations tailored toward them. Examples of these engines are Camunda\footref{footnote:camunda} or Bizagi\footref{footnote:bizagi}.

We acknowledge the work by Ding et al.~\cite{DBLP:conf/icnc/DingGM16} that, to the best of our knowledge, first proposed to transform procedural business process models into CEP rules. However, the work was focused on automating the decision points in the model rather than end-to-end orchestration. CEP rules have been used to execute declarative models or constraints for a long time~\cite{DBLP:conf/bis/RuhkampS21,DBLP:conf/sac/AwadBESAS15}

This work bears similarities with literature in hybrid process models (e.g. \cite{ALMAN2023102271,DBLP:conf/otm/SlaatsSMR16}). Alman et al. \cite{ALMAN2023102271} present an execution and monitoring semantics for declarative and imperative specifications using encodings on Data Petri Nets, and Slaats et al. introduce a refinement semantics where declarative and imperative processes can be merged via activity refinement~\cite{DBLP:conf/otm/SlaatsSMR16}. Our work aims at being more flexible than these attempts by only requiring notations to provide a semantics in terms of LTSs, so both encodings on Petri nets or based on programming languages can be supported.


\section{Conclusion and Future Work}
\label{sec:conclusion}

This paper introduced a novel approach for orchestrating business processes, which leverages streams, tables, and CQL to achieve its goal. The paper presents the rules to capture and implement BPMN and DCR processes in Esper, but other process modeling languages and complex event processing systems can be used.

It is important to point out that the unifying framework for process notations presented in this paper brings new questions to the state of the art in hybrid processes. By not relying on a common formalism, the conditions for soundness and compositionality of processes in the literature will need to be revised. In particular, our future work would like to study what are the minimal set of conditions for the composition of abstract process models that still preserve soundness properties such as termination and absence of deadlocks.

In the future, we plan to extend the set of rules to support other process modeling notations, thus allowing for processes expressed in hybrid notations (i.e., processing where parts of them are in one language and other parts are in another language) to be executed as well.

\balance

\bibliographystyle{plain}
\bibliography{ref}

\end{document}